\documentclass[prd,aps,showpacs,floats,letterpaper,floatfix,
groupedaddress,superscriptaddress
,nofootinbib
,eqsecnum,twocolumn
]{revtex4}
\usepackage{amsmath}
\usepackage{amsfonts}
\usepackage{bm}
\usepackage{graphicx}

\def\v1v2{{\bf v}_1 \cdot {\bf v}_2}
\allowdisplaybreaks

\begin{document}

\title{On incorporating post-Newtonian effects in $N$-body dynamics}

\author{
Clifford M.~Will} \email{cmw@physics.ufl.edu}
\affiliation{Department of Physics, University of Florida, Gainesville, Florida 32611, USA}
\affiliation{GReCO, Institut d'Astrophysique de Paris, CNRS,\\ 
Universit\'e Pierre et Marie Curie, 98 bis Bd. Arago, 75014 Paris, France}

\date{\today}

\begin{abstract}
The increasing role of general relativity in the dynamics of stellar systems with central massive black holes, in the generation of extreme mass-ratio inspirals and tidal disruption events, and in the evolution of hierarchical triple systems inspires a close examination of how post-Newtonian effects are incorporated into $N$-body dynamics.
The majority of approaches incorporate relativity by adding to the Newtonian $N$-body equations the standard two-body post-Newtonian terms for a given star around the black hole or for the close binary in a triple system.   We argue that, for calculating the evolution of such systems over timescales comparable to the relativistic pericenter advance timescale, it is essential to include ``cross terms'' in the equations of motion.  These are post-Newtonian terms in the equation of motion of a given body that represent a coupling between the potential of the central black hole and the potential due to other stars in the system.  For hierarchical triple systems, these are couplings between the potential of the inner binary and that of the distant third body.  Over pericenter precession timescales, the effects of such terms can actually be ``boosted'' to amplitudes of Newtonian order.   We write down the post-Newtonian $N$-body equations of motion including a central black hole in a truncated form that includes all the relevant cross terms, in a format ready to use for numerical implementation.   We do the same for hierarchical triple systems, and illustrate explicitly the effects of cross terms on the orbit-averaged equations of evolution for the orbit elements of the inner binary for  the special case where the third body is on a circular orbit.   We also describe in detail the inspiration for this investigation: the seemingly trivial problem of the motion of a test body about a central body with a Newtonian quadrupole moment, including the relativistic pericenter advance, whose correct solution for the conserved total Newtonian energy {\em requires} including post-Newtonian cross terms between the mass monopole potential and the quadrupole potential.

\end{abstract}

\maketitle

\section{Introduction and summary}
\label{sec:intro}

The effects of general relativity on the evolution of systems with more than two bodies have been studied intermittently for over 75 years.  Einstein~\cite{Einstein39} considered a crude model of a spherically symmetric star cluster where all the stars resided on circular orbits, in order to explore whether what we now call the event horizon could occur in nature.  The discovery of quasars in the early 1960s led to the suggestion~\cite{HoyleFowler67} that their large redshifts were due, not to cosmological expansion, but to the intrinsic redshift of light from the vicinity of a compact object.  One candidate was a highly relativistic stellar cluster.  However, detailed analyses revealed that they were subject to instabilities leading to catastrophic collapse at values of their central
redshifts that were far lower than those being measured for many quasars~\cite{1969ApJ...158...17I}.  In the end, the cosmological interpretation prevailed.  

In recent years, there has been renewed attention to the relevance of general relativity to multiple-star systems because of the observation that many galaxies, including our own, appear to harbor massive black holes in their cores (see~\cite{2005PhR...419...65A,2010RvMP...82.3121G,2013degn.book.....M} for reviews).  The capture or tidal disruption of stars passing close to such black holes could have important observational signatures in the electromagnetic spectrum~\cite{2013BrJPh.tmp...36G}.  Inspiralling relativistic orbits of stars or stellar-mass black holes around massive black holes (extreme mass-ratio inspirals, or EMRIs) or the inspiral of two massive black holes could generate gravitational radiation in the low-frequency regime detectable by a space-based laser interferometer (see \cite{2012arXiv1205.5240A} for a review).   Measurements of the precessing orbits of stars very close to the black hole SgrA* at the center of the Milky Way could yield tests of the ``no-hair'' theorems of general relativity~\cite{2008ApJ...674L..25W}.   While the examples just cited are essentially two-body problems, they cannot be discussed in complete isolation from the many-body context.  

Galactic cores contain many stellar objects (to say nothing of gas, dust and dark matter), and these objects have interacted and continue to interact with any two-body system that one might wish to study.  Whether tidal disruption events or stellar captures are frequent enough to be observable depends on whether interactions of a given star with the surrounding cluster have been sufficiently effective in diverting the star's orbit toward what is an extremely tiny target on galactic scales -- the black hole.  Likewise, whether a stellar-mass black hole can be aimed sufficiently close to the central massive black hole to allow gravitational radiation reaction to take over and induce an inspiral depends sensitively on the history of its interactions with the other stars and the black hole.   Whether two massive black holes are able to ``find'' each other following the merger of two galaxies that hosted each of them depends on their interactions with the other stars in the merged galaxy.  Finally, the ability to test general relativity at the galactic center depends on whether the cluster of stars that is likely to reside there perturbs the orbit of a given star so strongly as to swamp the relativistic precession effects being explored.

Generally speaking, stars are sufficiently far from the black hole that relativistic effects are much smaller than their mutual gravitational perturbations.  So what role does general relativity play in such an $N$-body context?   One recent example illustrates a potential role.

Repeated interactions of a given star in a dominantly Keplerian orbit about the black hole with the other stars in a cluster lead to torques that cause the eccentricity of the orbit to perform a quasi-random walk, while its orbital energy is roughly constant, a phenomenon known as resonant relaxation~\cite{1996NewA....1..149R}.  In some cases, this can lead to very large eccentricities, which can place the star on an orbit with a small enough pericenter relative to the central black hole that it is capable of losing energy to gravitational radiation and becoming an EMRI.   Hopman and Alexander\cite{2006ApJ...645.1152H} showed that the relativistic precession of the pericenter could act to suppress such torques.  In a series of numerical $N$-body simulations, Merritt et al.~\cite{mamw2}  studied this effect in detail.   The build-up of eccentricity of a given Keplerian orbit depends on the repeated coherent interactions with a set of stars whose orbits are in the ``vicinity'' of the eccentric orbit of the target star, and this build-up occurs over a long timescale.  However, the relativistic pericenter advance produces a non-random, secular precession of the orbit within its plane that renders these stellar torques ineffective.  This destroys the coherence that was driving the eccentricity toward high values, and leads to a cap on the eccentricity.  This occurs when the pericenter precession timescale, which decreases with increasing eccentricity, becomes comparable to the timescale for changes in angular momentum due to stellar torques.  Merritt et al. dubbed this the ``Schwarzschild barrier'', and showed that it had a significant effect on the rate of production of EMRI orbits.   The point was that, while relativistic effects are typically small, over a relativistic precession timescale, they can have significant effects in such $N$-body systems, because their timescales can be shorter than the other important timescales in the problem. 

In~\cite{mamw2}, general relativity was taken into account using the post-Newtonian (PN) approximation.  The first PN approximation is adequate for studying the overall evolution of such systems; higher PN effects, such as those due to gravitational radiation reaction, do not come into play until the star is well into its inspiral evolution.  Incorporating 1PN effects in an $N$-body evolution would appear to be straightforward -- simply replace Newtonian equations of motion with the post-Newtonian $N$-body equations of motion commonly known as the EIH equations, named for Einstein, Infeld and Hoffmann, who published them in 1938~\cite{EIH}, although equivalent equations were derived by Lorentz and Droste in 1917~\cite{LorentzDroste}.  However, these equations of motion contain 3-body interactions, and as the number $N$ of bodies grows, the computational burden of including these interactions becomes prohibitive.  An alternative approach might be to incorporate PN effects into the collisionless Boltzmann equation~\cite{2011PhRvD..83l3007A,2012PhRvD..86d3008R}, but this would be purely statistical in nature.   Accordingly, to our knowledge, all current $N$-body codes containing a central black hole include only the PN effects on each star due to the black hole, as if the star and black hole were a two-body system.

The purpose of this paper is to show that there is reason to believe that this may not be sufficient.  We will argue that, for problems involving evolutions over many relativistic precession timescales, it may necessary to include PN ``cross terms'' in the equations of motion in order to capture properly the relativistic effects of the black hole.  While the direct PN terms in the acceleration $\bm{a}_a$ of body $a$ caused by the black hole are proportional to $M^2$, where $M$ is the mass of the black hole, these cross terms are proportional to $M \times m_b$, where $m_b$ is the mass of one of the stars.  The remaining PN terms in $\bm{a}_a$ that are computationally troublesome, proportional to $m_b \times m_c$ are dropped.  

The basic idea is this:  in dimensionless terms, a relativistic effect induced by the black hole, such as the pericenter advance, is proportional to $GM/ac^2$, where $a$ is the semimajor axis of the body's orbit, and $G$ and $c$ are the gravitational constant and speed of light, respectively.   A Newtonian effect due to other stars is proportional to $(m_b/m_a)(a/r_{ab})^n$, where $n$ is some power which depends on the relative size of $a$ and $r_{ab}$ ($n$ is positive if $a < r_{ab}$, negative if $a > r_{ab}$).   A PN effect due to the ``cross terms'' described above would be proportional to $(GM/ac^2) \times (m_b/m_a)(a/r_{ab})^n$.  On the face of it, this is a smaller effect than either the pure black hole effect or the stellar effect, when $(a/r_{ab})^n \ll 1$.  However, if it is a secular effect, {\em and} if one is interested in how this effect grows over a relativistic timescale induced by the black hole, which is proportional to $1/(GM/ac^2)$, then the effect could be ``boosted'' from a PN level effect to a Newtonian level effect.  This could have hitherto unforeseen consequences in long-term evolutions of such systems.  

In this paper, we explicitly write down the truncated 1PN equations of motion for an $N$-body system with a central (non-rotating) black hole, including the cross terms to a consistent order, and display them in a ready-to-use form.  The equations include the appropriate equation of motion for the central black hole, and the appropriate expressions for globally conserved quantities such as energy, momentum and the center of mass.  These equations can be used in numerical experiments to explore the impact of including PN cross terms.

Another system where relativistic effects have been studied is the hierarchical three-body system, where a close binary system is in orbit with a distant third body.  In Newtonian theory, such systems exhibit the remarkable Kozai-Lidov effect, in which there is an interchange between the eccentricity of the two-body orbit and its inclination relative to the plane of the third body.  Under certain circumstances, the system can exhibit the Kozai resonance, in which the pericenter of the two-body orbit oscillates about either $\pi/2$ or $3\pi/2$, while the eccentricity and inclination oscillate about values related by $5\cos^2 \iota = 3(1-e^2)$.   However, the pericenter precession of the two-body orbit induced by general relativity can force the system out of the resonant state.  Here again is a problem which mixes a Newtonian effect and a relativistic effect, in which the relativistic precession timescale is relevant.  Could PN cross terms play a role here?  In this case the dimensionless size of the expected cross terms is proportional to $(Gm/ac^2) \times (m_3/m)(a/R)^3$, where $m$ is the total mass of the two-body system, and $R$ is the distance of the third body.

Efforts to include relativity in Kozai-Lidov systems have taken three approaches.  One is to include the standard PN terms in the dynamics of the inner binary, to an order as high as 2.5PN order, but without cross terms~\cite{2002ApJ...576..894M,2002ApJ...578..775B,2003ApJ...598..419W,2011MNRAS.411..565M,2013PhRvL.111f1106S,2013arXiv1308.5682A}.  Here it is possible to solve the orbit perturbation equations in the same manner as in the Newtonian Kozai problem,  by averaging over an orbit of both the inner system and the third body.  The result is a set of average equations for the evolution of the orbit elements whose long-term evolution can be studied numerically.   

Another approach~\cite{2013ApJ...773..187N} begins with the full EIH three-body PN equations of motion, frequently expressed in Hamiltonian form.   After expressing the Hamiltonian in Delaunay variables and averaging over both inner and outer orbits, one obtains average equations of evolution for the orbit elements.  While this approach automatically incorporates all PN terms initially, we shall see in the next section that there are subtleties in the averaging procedure that may have been overlooked.  

The third approach carries out a full numerical integration of the complete three-body PN equations of motion.  In \cite{2011PhRvD..83h4013G,2011PhRvD..84j4038G}, the equations were in fact complete and correct through 2.5PN order.   While this approach in principle captures all the relevant effects, it requires performing many numerical experiments and carefully analysing the output data.  As complete as this approach may be, it does not easily lend itself to exploring the underlying physical phenomena. 

In this paper, we write down the truncated post-Newtonian equations of motion for hierarchical triple systems including the relevant cross terms for both the two-body system and the third body relative to the two-body center of mass.  We then carry out the standard double average of the orbit perturbation equations for the special case of a circular third-body orbit (with an important caveat, to be discussed below), finding averaged equations for the orbit elements of the two-body system that are amenable to numerical integration.  Exploration of the impact of PN cross terms on these averaged equations will be left to future work.

We begin in Sec.\ \ref{sec:conundrum} by describing the simple physical problem that inspired these considerations: the motion of a body in the field of a central object with a mass and a quadrupole moment, including PN effects.  This seemingly trivial problem was the origin of a conundrum encountered during the work leading to Ref.~\cite{mamw2}.  The resolution of this conundrum absolutely required the incorporation of the relevant PN cross terms.  In Sec.\ \ref{sec:PNNbody} we display the truncated equations of motion for an $N$-body system and a central black hole, along with the relevant conserved quantities.  Section \ref{sec:triples} displays the equations for hierarchical three-body systems and derives the orbit averaged equations for the evolution of the orbit elements, including all cross-term effects.  Concluding remarks are made in Sec.\ \ref{sec:conclusions}.

\section{An object lesson: The quadrupole conundrum}
\label{sec:conundrum}

The origin of the idea that it may be important to incorporate appropriate PN cross terms in long-term evolutions of orbital dynamics was a series of numerical simulations of $N$-body orbits with a central massive rotating black hole carried out by Merritt et al.~\cite{mamw2}.  One set of long-term numerical integrations involved single-particle orbits around a rotating black hole, with the relativistic effects due to the hole expanded in the post-Newtonian approximation.  In fact the essential effects that lead to the conundrum of the title of this section were seen by turning off the frame-dragging terms, and keeping only the quadrupole moment of the black hole.  The orbit of the particle was chosen to have a large eccentricity, and the integrations were carried out over several precessions of the orbital pericenter $\omega$, induced dominantly by the standard ``point-mass'' pericenter advance of $6\pi GM/a(1-e^2)c^2$ per orbit.  

It was found that the semi-major axis $a$ of the orbit varied as $\sin^2 \omega$, was proportional to $\sin^2 \iota$, where $\iota$ is the inclination of the orbit relative to the equatorial plane of the black hole, and had an anomalously large amplitude, so large as to suggest potentially important observable effects in $N$-body systems orbiting a rotating black hole with a quadrupole moment.    

Physically, there is nothing mysterious about a variation of $a$ with $\omega$.  The Newtonian conserved energy per unit mass of a body orbiting a mass $M$ with a Newtonian quadrupole moment $Q_2$ is given by
\begin{equation}
E = - \frac{GM}{2a} - \frac{GQ_2}{2}\left ( \frac{1+e}{p} \right )^3 \bigl (3\sin^2\iota\sin^2 \omega -1 \bigr) \,,
\label{EnewtQ}
\end{equation}
where $p = a(1-e^2)$ is the semi-latus rectum.  Because the osculating orbit elements are in general functions of orbital phase in any non-Keplerian motion, they are defined in this example at pericenter. 
When $\omega =0$, the long axis of the orbit lies along the line of nodes, the intersection of the orbit plane with the equatorial plane of the central body, and both the pericenter and apocenter of the orbit occur on the equator of the central body.  For a highly eccentric orbit, the orbit is displaced only a small distance above or below the equatorial plane.   But when $\omega=\pi/2$, the long axis is perpendicular to the nodal line, and the pericenter occurs over one of the hemispheres, where the body sees a very different potential because of the non-zero $Q_2$, while the apocenter occurs far away in the opposite hemisphere.  Thus the quadrupole interaction energy is very different between the two cases.   Therefore, in order to keep $E$ constant while $\omega$ varies, $a$ must vary to compensate for the varying quadrupole interaction energy.   In fact it is easy to see from Eq.\ (\ref{EnewtQ}) that $a$ must vary, to first order in $Q_2/Ma_0^2$, according to  
\begin{equation}
a = a_0 + 3\frac{Q_2}{Ma_0 (1-e_0)^3} \sin^2 \iota_0 
 \sin^2 \omega \,,
 \label{avariation0}
\end{equation}
where the subscripts denote values of the orbit elements at an initial value of $\omega$.

However, this result presents us with two conundra:
\begin{itemize}
\item
In first-order perturbation theory, for motion in the field of a point mass, with point-mass 1PN corrections, plus a Newtonian quadrupole term, the net change in $a$ over one orbit vanishes identically, so what is the origin of this variation over a precession timescale?
\item
The amplitude of variation in $a$ that emerges from this argument disagrees with the amplitude that emerged from the numerical simulations in~\cite{mamw2}.
\end{itemize}

A resolution to this conundrum is suggested by the following fallacious argument.  Using the first-order orbital perturbation equations, evaluate the change in $a$ not after a complete orbit (a change $2\pi$ in the true anomaly $f$ of the orbit), where the change vanishes, but after one orbit plus the small advance of pericenter $6\pi GM/pc^2$.  This gives a change in $a$ that actually has the correct dependence on the variables.  It is fallacious because $a$ remains strictly periodic, with no net variation over a pericenter precession timescale.   But it gives a fractional change in $a$ over one ``orbit'' that is proportional to the product $(GM/pc^2) \times (Q_2/Mp^2)$, which represents a cross-term between the quadrupole and relativistic perturbations.  This suggests that, in order to find the full solution to the problem of the long-term evolution of $a$ with quadrupole and PN effects together, one must include the cross terms in the equations of motion.   In this section, we verify that this is correct with a straightforward analytical calculation.

We begin with the 1PN equation of motion
\begin{equation}
\frac{d{\bm v}}{dt} = {\bm \nabla} U + \frac{1}{c^2}{\bm \nabla} U \left ( v^2 - 4U \right) - \frac{4}{c^2}{\bm v} {\bm v} \cdot {\bm \nabla} U  \,,
\end{equation}
where the gravitational potential contains both a point mass and a quadrupole term,
\begin{equation}
U = \frac{GM}{r} + \frac{1}{2} \frac{GQ_2}{r^3} \left [ 3 ({\bm e} \cdot {\bm n} )^2 -1 \right ] \,,
\end{equation}
where $M$ and $Q_2$ are the mass and quadrupole moment of the source, ${\bm e}$ is a unit vector along the source's symmetry axis, ${\bm n} = {\bm x}/r$, and ${\bm v}$ is the velocity of the particle.   

These equations admit the conserved orbital energy and conserved ${\bm e}$-component of angular momentum (per unit mass)
\begin{subequations}
\begin{eqnarray}
E &=& \frac{1}{2} v^2 - U + \frac{1}{c^2} \left ( \frac{3}{8} v^4 + \frac{3}{2} v^2 U + \frac{1}{2} U^2 \right ) ,
\label{energy}
\\
h_e &\equiv& {\bm h} \cdot {\bm e} = r(v^2 - \dot{r}^2)^{1/2} \cos \iota \left ( 1 + \frac{4}{c^2} U \right ) ,
\label{angularmomentum}
\end{eqnarray}
\end{subequations}
where $\iota$ is the orbital inclination.

Then, to 1PN order, and including the Newtonian quadrupole {\em and} the quadrupole-PN cross terms, the equations of motion take the form
\begin{eqnarray}
\frac{d{\bm v}}{dt} &=& - \frac{GM{\bm n}}{r^2} \left [1 + \frac{1}{c^2} \left ( v^2 - 4\frac{GM}{r} \right ) \right ] + 4\frac{GM}{c^2 r^2} \dot{r} {\bm v}
\nonumber \\
&& 
- \frac{3}{2} \frac{GQ_2}{r^4} \left [ 5{\bm n}({\bm e} \cdot {\bm n})^2 - 2{\bm e}({\bm e} \cdot {\bm n}) - {\bm n} \right ] 
\nonumber \\
&&
-\frac{3}{2} \frac{GQ_2}{c^2 r^4} \left [ 5{\bm n}({\bm e} \cdot {\bm n})^2 - 2{\bm e}({\bm e} \cdot {\bm n}) - {\bm n} \right ] 
\nonumber \\
&&
\qquad \times \left (v^2 - \frac{4GM}{r} \right )
\nonumber \\
&&
+  \frac{6GQ_2}{c^2 r^4} {\bm v}  \left [ 5\dot{r}({\bm e} \cdot {\bm n})^2 - 2({\bm v} \cdot {\bm e})({\bm e} \cdot {\bm n}) - \dot{r} \right ] 
\nonumber \\
&&
+ \frac{2G^2 M Q_2}{c^2r^5} {\bm n} \left [3({\bm e} \cdot {\bm n})^2 -1 \right ] \,.
\label{eom_mixed}
\end{eqnarray}
The first line of Eq.\ (\ref{eom_mixed}) shows the Newtonian and PN point-mass terms, the second line shows the Newtonian quadrupole term, and the remaining lines show the mixed PN-quadrupole terms; it is useful to recall that $v^2 \sim GM/r$.    Terms proportional to $Q_2^2$ have been dropped.

We use standard orbital perturbation theory to compute the secular changes in the semi-latus rectum $p$, eccentricity $e$, inclination $\iota$, nodal angle $\Omega$ and pericenter angle $\omega$.   
For a general orbit, the osculating Keplerian orbit is defined by the following set of equations:
\begin{eqnarray}
r &\equiv& p/(1+e \cos f) \,,
\nonumber \\
{\bm x} &\equiv& r {\bm n} \,,
\nonumber \\
{\bm n} &\equiv& \left [ \cos \Omega \cos(\omega + f) - \cos \iota \sin \Omega \sin (\omega + f) \right ] {\bm e}_X 
\nonumber \\
&&
 + \left [ \sin \Omega \cos (\omega + f) + \cos \iota \cos \Omega \sin(\omega + f) \right ]{\bm e}_Y
\nonumber \\
&&
+ \sin \iota \sin(\omega + f) {\bm e}_Z \,,
\nonumber \\
{\bm \lambda} &\equiv& d{\bm n}/df \,, \quad \hat{\bm h}={\bm n} \times {\bm \lambda} \,,
\nonumber \\
{\bm h} &\equiv& {\bm x} \times {\bm v} \equiv \sqrt{GMp} \, \bm{\hat{h}} \,,
\label{keplerorbit}
\end{eqnarray}
where $f$ is the orbital phase, or {\em true anomaly}, and 
where ${\bm e}_A$ are chosen reference basis vectors, with ${\bm e}_Z$ parallel to the symmetry axis of the central mass.   From the given definitions, it is evident that ${\bm v} = \dot{r} {\bm n} + (h/r) {\bm \lambda}$ and $\dot{r} = (he/p) \sin f$.   One then defines the radial $\cal R$, cross-track $\cal S$ and out-of-plane $\cal W$ components of the perturbing acceleration $\delta {\bm a}$, defined respectively by
${\cal R} \equiv {\bm n} \cdot \delta {\bm a}$,
 ${\cal S} \equiv {\bm \lambda} \cdot \delta {\bm a}$ and
 ${\cal W} \equiv \bm{\hat{h}} \cdot \delta {\bm a}$,
and writes down the ``Lagrange planetary equations'' for the evolution of the orbit elements,
\begin{eqnarray}
\frac{dp}{dt} &=& 2 \sqrt{\frac{p^3}{GM}} \frac{{\cal S}}{1+e \cos f} \,,
\nonumber \\
\frac{de}{dt} &=& \sqrt{\frac{p}{GM}} \left [ \sin f \, {\cal R} + \frac{2\cos f + e +e\cos^2 f}{1+ e\cos f} {\cal S} \right ]\,,
\nonumber \\
\frac{d\omega}{dt} &=& \frac{1}{e}\sqrt{\frac{p}{GM}} \left [ -\cos f \, {\cal R} + \frac{2 + e\cos f}{1+ e\cos f}\sin f {\cal S} 
\right .
\nonumber \\
&&
\left . \quad - e\cot \iota \frac{\sin (\omega +f)}{1+ e\cos f} {\cal W} \right ] \,,
\nonumber \\
\frac{d\iota}{dt} &=& \sqrt{\frac{p}{GM}} \frac{\cos (\omega +f)}{1+ e\cos f} {\cal W} \,,
\nonumber \\
\sin \iota \frac{d\Omega}{dt} &=& \sqrt{\frac{p}{GM}} \frac{\sin (\omega +f)}{1+ e\cos f} {\cal W} \,.
\end{eqnarray}
To first order in either the PN or quadrupole perturbations, we can substitute $d/dt = (h/r^2) d/df$ and integrate these equations over $f$ between $0$ and $2\pi$, holding the orbit elements on the right-hand-side fixed to their initial values.  The result is $\Delta p = 0$, $\Delta e =0$, $\Delta \iota = 0$,  $\Delta \Omega = 3\pi (Q_2/Mp^2) \cos \iota$, and $\Delta \omega = 6\pi GM/c^2p - 6\pi (Q_2/Mp^2) (1-5 \sin^2 \iota /4)$.

In order to find the changes in the orbit elements induced by the PN-quadrupole cross terms in the equations of motion, we must incorporate higher-order effects in the perturbation equations themselves, as follows:

\begin{enumerate}

\item
We must carry out the orbital perturbation theory carefully.  The orbit elements $a$, $e$, $i$, $\Omega$ and $\omega$ vary periodically during the orbit.  Thus the PN-induced variations in these elements must be inserted back into the perturbation terms generated by the quadrupole moment, and the quadrupole-induced variations must be inserted back into the perturbation terms generated by PN effects.  These will produce cross-term contributions of the same order as those from the equations of motion.
\item
In converting from $d/dt$ in the perturbation equations to $d/df$,  we must not use $df/dt = h/r^2$, but instead must use
\begin{equation}
\frac{df}{dt} = \frac{h}{r^2}  - \dot{\omega} - \dot{\Omega} \cos \iota \,.
\label{dfdt}
\end{equation}
The added terms come from the fact that, while $t$ is measured from a fixed moment of time, $f$ is measured from the pericenter, which changes via both $\dot{\omega}$ and $\dot{\Omega}$.    These terms will also generate cross-term effects between PN and quadrupole terms.  
\end{enumerate}

Choosing a fiducial true anomaly $f_0$ as the orbital phase where the initial values of the orbit elements are prescribed, we integrate  the orbit perturbation equations over $f$ from $f_0$ to $F$, to obtain the instantaneous values of the elements.   The expressions are too complicated to display, but we use them to verify that the energy is independent of true anomaly $F$, to order $G^2MQ^2/p^4c^2$, given by
\begin{widetext}
\begin{eqnarray}
E &=& - \frac{GM}{2a} - \frac{GQ_2}{2}\left ( \frac{1+e\cos f_0}{p} \right )^3 \bigl (3\sin^2 \iota \sin^2 (\omega + f_0) -1 \bigr) 
\nonumber \\
&& + \frac{1}{c^2} \biggl [ \frac{3}{8} \left (\frac{GM}{a} \right )^2
+ \left (\frac{GM}{p} \right )^2 (1+e \cos f_0)(2 + 5e\cos f_0 + 3e^2) 
\nonumber \\
&& - \frac{1}{4} \frac{G^2MQ_2}{p r_0^3}  (5 + 8e\cos f_0 + 3e^2) \bigl(3\sin^2\iota\sin^2 (\omega + f_0) -1\bigr )  \biggr ] \,,
\label{EtrueF}
\end{eqnarray}
where
\begin{equation}
r_0 = \frac{p}{1+e \cos f_0} \,,
\end{equation}
and all orbit elements are evaluated at $f_0$.  We also verify that $h_e$ is independent of $F$, through the analogous order.

Choosing $F=f_0 + 2\pi$, we then find expressions for the secular
variation of $p$, $e$ and $i$ over one complete orbit:
\begin{subequations}
\begin{eqnarray}
\Delta p &=& 12\pi  \frac{GQ_2}{p^2c^2} \sin^2\iota\, P'\,,
\label{deltaa}
\\
\Delta e &=& -3\pi\frac{GQ_2}{p^3c^2}  \sin^2\iota\, \left [ 3 (1+ e\cos f_0)^3 \sin 2(\omega + f_0)
+ 2 (1-e^2) P' \right ]
\,,
\label{deltae}
\\
\Delta\iota&=& 6\pi \frac{GQ_2}{p^3c^2}  \sin\iota\cos\iota\,P' \,,
\label{deltai}
\end{eqnarray}
\label{deltaelements}
\end{subequations}
where $P$ is a function of $e_0$, $f_0$ and $\omega$, given by
\begin{eqnarray}
P(e_0, f_0, \omega ) &\equiv& (e_0+3\cos f_0 + 2e_0\cos^2 f_0) \sin \omega \sin(\omega + f_0) 
\nonumber \\
&& \quad - \frac{1}{4} (6 + 4e_0\cos f_0 - 3e_0^2 ) \sin^2 \omega \,,
\end{eqnarray}
with $P' \equiv \partial P/\partial \omega$.
The presence of both $Q_2$ and $c^2$ indicates that these are PN--cross-term effects.
These are the secular changes in $p$, $e$ and $\iota$ over one orbit.  Nominally they would grow linearly in time, except for the fact that the angle of pericenter $\omega$ is changing with time at the rate per orbit dominated by the Schwarzschild mass term, namely $\Delta \omega = 6\pi GM/pc^2$.  Thus we can combine this with Eqs.\ (\ref{deltaelements}),  to obtain the equation, for each element $Z$, 
\begin{equation}
Z = Z_0 + \int (\Delta Z/\Delta t) dt = Z_0 + \int (\Delta Z/\Delta \omega) d\omega \,. \label{integrate_element}
\end{equation}
Given that $p$, $e$ and $\iota$ are constant to lowest order at this level of approximation, these equations integrate to the expressions,
\begin{subequations}
\begin{eqnarray}
p&=& p_0 + \frac{2Q_2}{Mp_0} \sin^2 \iota_0 \,  P \,,
\label{psol}
\\
e &=&e_0 + \frac{1}{2}  \frac{Q_2}{Me_0p_0^2} \sin^2 \iota_0 
\biggl [ 3(1+e_0 \cos f_0 )^3 (\sin^2 (\omega + f_0) - \sin^2 f_0 )
 - 2 (1-e_0^2) P \biggr ] \,,
\label{esol}
\\
\iota &=& \iota_0 +  \frac{Q_2}{Mp_0^2}
\sin \iota_0 \cos \iota_0 \,  P \,,
\label{isol}
\end{eqnarray}
\label{peisol}
\end{subequations}
where $p_0$, $e_0$, $\iota_0$ are to be evaluated at $f_0$.  Notice that $c^{-2}$ has disappeared from the expressions, so that the perturbations over the precession timescale are now of a Newtonian quadrupole order.
Combining the results for $p$ and $e$, we obtain for the semi-major axis,
\begin{equation}
a = a_0 + 3\frac{Q_2}{Mp_0 (1-e_0^2)^2} \sin^2 \iota_0 
(1+e_0 \cos f_0 )^3 (\sin^2 (\omega + f_0) - \sin^2 f_0 ) \,.
\end{equation}
The variations in the orbital elements depend on the fiducial phase $f_0$ because the orbital elements vary over an orbital timescale.  Thus the orbit elements $a$, $e$, $\iota$, $\omega$ and $\Omega$ inferred from a given set of initial conditions ${\bm x}_0$ and ${\bm v}_0$ depend on the choice of $f_0$, which here is a proxy for the sixth orbit element, the time of pericenter passage.  For $f_0 = 0$, the variation in $a$ agrees with Eq.\ (\ref{avariation0})

Notice that, with this variation of $a$ over a pericenter advance timescale, the Newtonian and quadrupole energy together in Eq.\ (\ref{EtrueF}) are independent of $\omega$, taking the form
\begin{equation}
E = - \frac{GM}{2a_0} - \frac{GQ_2}{2}\left ( \frac{1+e\cos f_0}{p_0} \right )^3 \bigl (3\sin^2\iota\sin^2 f_0 -1 \bigr) \,.
\end{equation}
It was {\em essential} to find the post-Newtonian solution for $\Delta p$, $\Delta e$, $\Delta i$ and $\Delta a$ consistently to $O(GQ_2/c^2)$ in order to conserve the energy to {\em Newtonian} quadrupole order over the pericenter advance timescale.   Basically the secular PN pericenter advance promotes the PN-quadrupole cross-term perturbations to a lower order of approximation.
It was also essential to carry out the orbital perturbation theory carefully, following points 1 and 2 above; a failure to do so would not have given a variation in $a$ that would lead to a conserved energy.   A number of earlier works that included the cross-terms in the equations of motion failed to heed those two points~\cite{1988CeMec..42...81S,1991ercm.book.....B} and thus did not obtain the correct long-term evolution of the orbit elements. 

Because $a$ and $e$ vary with $\omega$, the inclination must also vary, in order to keep $h_e$ constant.  This is a kind of Kozai mechanism, induced again by the relativistic precession of $\omega$. 

The contributions to the energy proportional to $G^2MQ_2/p^4$ in Eq.\ (\ref{EtrueF}), while independent of $F$,  cannot be shown to be independent of $\omega$; to do so would require a calculation of the orbit-element evolutions through $O(G^3 M^2 Q_2/p^5)$.  

For the special case of $f_0 = \pi$, the variations in the orbit elements are given by
\begin{subequations}
\begin{eqnarray}
p&=& p_0 + \frac{Q_2}{2Mp_0} (6-8e_0+3e_0^2)\sin^2 \iota_0 \, \sin^2 \omega  \,,
\label{psol2}
\\
e &=&e_0 - \frac{1}{4}  \frac{Q_2}{Mp_0^2} (2-e_0)(5-3e_0)(1-e_0)\sin^2 \iota_0 \, \sin^2\omega \,,
\label{esol2}
\\
\iota  &=& \iota_0 + \frac{1}{4} \frac{Q_2}{Mp_0^2}  (6-8e_0+3e_0^2)
\sin \iota_0 \cos \iota_0 \, \sin^2 \omega \,,
\label{isol2}
\\
a&=& a_0 + 3\frac{Q_2}{Mp_0} \frac{1-e_0}{(1+e_0)^2} \sin^2 \iota_0 \, \sin^2 \omega \,.
\label{asol2}
\end{eqnarray}
\label{peisol2}
\end{subequations}
\end{widetext}

The variation of $a$ in Eq.\ (\ref{asol2}) still does not agree with the large amplitude variation found in the numerical evolutions carried out in the course of the work by Merritt {\em et al.}\cite{mamw2}.   (In those simulations, the orbit elements were inferred from the particle's position and velocity at apocenter, $f_0 = \pi$.)  The numerical codes did yield a variation in $a$ over a pericenter precession timescale, notwithstanding the fact that first-order perturbation theory yields nothing, because, being exact integrations of the equations of motion, they inherently incorporated the higher-order effects discussed in points 1 and 2 above.  However, it turns out that the codes used in those simulations did {\em not} initially contain the direct PN-quadrupole terms of Eq.\ (\ref{eom_mixed}).  When those terms were added to the codes, the integrations yielded variations in $a$ in {\em perfect} agreement with Eq.\ (\ref{asol2}).    
 
The lesson learned from this is that, when carrying out integrations of the dynamics of bodies  with relativistic effects over timescales comparable to the pericenter precession time, it is necessary to include cross terms between the relativistic perturbations and the perturbations from the other source, whether it be a quadrupole perturbation, or a perturbation due to one or more distant bodies.    It is to this latter context that we now turn.

\section{Post-Newtonian effects of a central black hole in $N$-body dynamics}
\label{sec:PNNbody}

An important class of $N$-body problems where relativity may play a role involves stellar clusters around massive black holes.  From a relativistic point of view, a naive approach to such problems would be simply to apply the post-Newtonian N-body equations of motion, which have been known since the 1917 work of Lorentz and Droste~\cite{LorentzDroste}, but are more commonly referred to as the Einstein-Infeld-Hoffman (EIH) equations of motion\cite{EIH}.   In these equations, the acceleration of body $a$ is given by 
\begin{widetext}  
\begin{eqnarray}
{\bm a}_a &=&  -\sum_{b \ne a} \frac{Gm_b {\bm x}_{ab}}{r_{ab}^3}
\nonumber \\
&&  + \frac{1}{c^2} \sum_{b \ne a} \frac{Gm_b {\bm x}_{ab}}{r_{ab}^3}
\biggl [ 4 \frac{Gm_b}{r_{ab}} + 5\frac{Gm_a}{r_{ab}} +
\sum_{c \ne a,b} \frac{Gm_c}{r_{bc}} 
+ 4 \sum_{c \ne a,b} \frac{Gm_c}{r_{ac}}
\nonumber \\
&& \qquad 
- \frac{1}{2} \sum_{c \ne a,b} \frac{Gm_c}{r_{bc}^3} (\bm{x}_{ab} \cdot \bm{x}_{bc}) - v_a^2 + 4 \bm{v}_a \cdot \bm{v}_b - 2 \bm{v}_b^2 + \frac{3}{2} (\bm{v}_b \cdot \bm{n}_{ab} )^2 \biggr ]
\nonumber \\
&& 
-\frac{7}{2c^2}   \sum_{b \ne a} \frac{Gm_b}{r_{ab}} \sum_{c \ne a,b} \frac{Gm_c \bm{x}_{bc}}{r_{bc}^3}  
+ \frac{1}{c^2} \sum_{b \ne a} \frac{Gm_b}{r_{ab}^3}  {\bm x}_{ab} \cdot (4 \bm{v}_a - 3 \bm{v}_b ) (\bm{v}_a - \bm{v}_b ) \,,
\label{eq:EIH}
\end{eqnarray}
where $\bm{n}_{ab} = \bm{x}_{ab}/r_{ab}$.  
These equations admit 1PN-accurate conserved orbital energy and linear momentum, given by
\begin{subequations}
\begin{eqnarray}
E  &=& \frac{1}{2} \sum_{a} m_a \left ( v_a^2 - \sum_{b \ne a} \frac{Gm_b}{r_{ab}} \right )
+ \frac{1}{c^2} \sum_{a} m_a \biggl [ \frac{3}{8} v_a^4 + \frac{3}{2} v_a^2 \sum_{b \ne a} \frac{Gm_b}{r_{ab}}  + \frac{1}{2} \sum_{b \ne a} \sum_{c \ne a} \frac{G^2 m_b m_c}{r_{ab} r_{ac}} 
\nonumber \\
&& \qquad \qquad - \frac{1}{4} \sum_{b \ne a} \frac{Gm_b}{r_{ab}} \left ( 7 {\bm v}_a \cdot {\bm v}_b
+ (\bm{v}_a \cdot \bm{n}_{ab})(\bm{v}_b \cdot \bm{n}_{ab}) \right ) 
\biggr ]
 \,,
\label{energy2}
\\
\bm{P} &=& \sum_{a} m_a \bm{v}_a 
+ \frac{1}{2c^2} \sum_{a} m_a \bm{v}_a 
\left ( v_a^2 - \sum_{b \ne a} \frac{Gm_b}{r_{ab}} \right )
- \frac{G}{2c^2} \sum_a \sum_{b \ne a} \frac{m_a m_b}{r_{ab}} (\bm{v}_a \cdot \bm{n}_{ab}) \bm{n}_{ab}  \,.
\label{momentum}
\end{eqnarray}
\label{energymomentum}
\end{subequations}

However, to incorporate these fully 1PN-accurate equations would not be practical, at least for large numbers of particles, because of the computational burden of incorporating the three-body interactions represented by the terms involving sums over $c \ne a,\, b$.   However, as the discussion of the previous section indicates, we are really interested only in the cross terms between the mass of the central black hole and the mass of a given star, not in the PN star-star terms.

Accordingly we truncate the EIH equations of motion, keeping only the Newtonian terms from the black hole and the $N$-bodies, plus the 1PN terms from the black hole alone, i.e. terms proportional to $G^2 M^2/r_{1a}^3c^2$, or  to $(GM/r_{1a}) (v_a^2 /c^2)$, where $M$ is the mass of the black hole, plus 1PN terms of the schematic form $G^2Mm_a/r^3c^2$, where $r$ here represents various interparticle distances.  
Note that $v_a^2 \sim GM/r_{1a}$, while $v_1 \sim (m_a/M) v_a$, so we use the same argument when considering velocity-dependent terms.

We denote the black hole by body \#1, and we take into account the conservation of linear momentum at Newtonian order, $M{\bm v}_1 + \sum_a m_a {\bm v}_a =0$, in order to eliminate ${\bm v}_1$ from post-Newtonian terms in the equations of motion.  
The resulting equations of motion have the form
\begin{equation}
{\bm a}_a = - \frac{GM{\bm x}_{a1}}{r_{a1}^3} - \sum_b \frac{Gm_b{\bm x}_{ab}}{r_{ab}^3}
+ \frac{1}{c^2}[{\bm a}_a]_{\rm BH} + \frac{1}{c^2}[{\bm a}_a]_{\rm Cross}
+ O\left (\frac{G^2 m_b^2}{c^2 r^3}\right ) \,,
\label{acc1}
\end{equation}
where
\begin{subequations}
\begin{eqnarray}
[{\bm a}_a]_{\rm BH} &=& \frac{GM{\bm x}_{a1}}{r_{a1}^3} \left ( 4\frac{GM}{r_{a1}} - v_a^2 \right ) + 4\frac{GM}{r_{a1}^3} \left ({\bm v}_a \cdot {\bm x}_{a1} \right ) {\bm v}_a  \,,
\label{acc1PN}
\\
{[{\bm a}_a]}_{\rm Cross} 
&=& 5\frac{G^2 m_aM {\bm x}_{a1}}{r_{a1}^4}
- \frac{G m_a }{r_{a1}^3} \left [ 4 v_a^2 {\bm x}_{a1} - 7( {\bm v}_a \cdot {\bm x}_{a1}){\bm v}_a \right ]
\nonumber \\
&& 
+\sum_b \frac{G^2 m_bM {\bm x}_{a1}}{r_{a1}^3} \left ( \frac{4}{r_{ab}} + \frac{5}{4r_{b1}} + \frac{r_{a1}^2}{4r_{b1}^3}  -\frac{r_{ab}^2}{4r_{b1}^3} \right )
\nonumber \\
&& 
+ \sum_b \frac{G^2 m_bM {\bm x}_{ab}}{r_{ab}^3} \left ( \frac{4}{r_{a1}} + \frac{5}{4r_{b1}} - \frac{r_{a1}^2}{4r_{b1}^3}  +\frac{r_{ab}^2}{4r_{b1}^3} \right )
\nonumber \\
&&
- \frac{7}{2} \sum_b \frac{G^2 m_bM {\bm x}_{b1}}{r_{b1}^3} \left ( \frac{1}{r_{ab}} - \frac{1}{r_{a1}} \right )
\nonumber \\
&&
- \sum_b  \frac{G m_b }{r_{a1}^3}  \left [ 4 ({\bm v}_a \cdot {\bm v}_b ){\bm x}_{a1} -3 ({\bm v}_b \cdot  {\bm x}_{a1} ) {\bm v}_a - 4 ({\bm v}_a \cdot  {\bm x}_{a1} ) {\bm v}_b \right ]
\nonumber \\
&&
+\sum_b \frac{G m_b {\bm x}_{ab}}{r_{ab}^3} \left [ v_a^2 - 2 |{\bm v}_{ab}|^2 + \frac{3}{2} \left ( {\bm v}_b \cdot {\bm n}_{ab} \right )^2 \right ] 
\nonumber \\
&&
+\sum_b \frac{G m_b }{r_{ab}^3} \left [ {\bm x}_{ab} \cdot (4{\bm v}_a -3{\bm v}_b ) \right ]{\bm v}_{ab} \,,
\label{acc1cross}
\end{eqnarray}
\label{acc1combined}
\end{subequations}
where ${\bm v}_{ab} \equiv {\bm v}_a - {\bm v}_b$, and the sum over $b$ excludes both body $a$ and the black hole.  We are basically ignoring PN terms that involve only the star-star interactions; in this truncated form the equations contain only two-body interactions, no worse computationally than the original Newtonian equations plus the PN terms from the central black hole.

The equation of motion for the black hole to the equivalent order of approximation is given by 
\begin{equation}
{\bm a}_1 = - \sum_b \frac{Gm_b{\bm x}_{1b}}{r_{1b}^3}
+ \frac{1}{c^2}[{\bm a}_1]_{\rm BH} + \frac{1}{c^2}[{\bm a}_1]_{\rm Cross} + 
O\left ( \frac{G^2 m_a^3}{Mc^2r^3} \right ) \,,
\label{accH}
\end{equation}
where
\begin{subequations}
\begin{eqnarray}
[{\bm a}_1]_{\rm BH} &=& \sum_b \frac{Gm_b{\bm x}_{1b}}{r_{1b}^3} \left ( 5\frac{GM}{r_{1b}} - 2 v_b^2 + \frac{3}{2} ({\bm v}_b \cdot {\bm n}_{1b} )^2 \right ) 
\nonumber \\
&&
+ 3\sum_b \frac{Gm_b}{r_{1b}^3} \left ({\bm v}_b \cdot {\bm x}_{1b} \right ) {\bm v}_b  \,,
\label{accHPN}
\\
{[{\bm a}_1]}_{\rm Cross} 
&=& 4\sum_b \frac{G^2m_b^2{\bm x}_{1b}}{r_{1b}^4} 
\nonumber \\
&&
+ \sum_{b,c} \frac{G^2m_b m_c {\bm x}_{1b}}{r_{1b}^3} \left ( \frac{4}{r_{1c}} + \frac{5}{4r_{bc}} - \frac{r_{1c}^2}{4r_{bc}^3} + \frac{r_{1b}^2}{4r_{bc}^3} \right )
\nonumber \\
&&
-\frac{7}{2} \sum_{b,c} \frac{G^2 m_b m_c {\bm x}_{bc}}{r_{bc}^3 r_{1b}}
\nonumber \\
&&
- \sum_{b,c} \frac{Gm_b m_c}{Mr_{1b}^3}  \left [ 4 ({\bm v}_b \cdot {\bm v}_c ){\bm x}_{1b} -3 ({\bm v}_b \cdot  {\bm x}_{1b} ) {\bm v}_c - 4 ({\bm v}_c \cdot  {\bm x}_{1b} ) {\bm v}_b \right ] \,.
\label{accHcross}
\end{eqnarray} 
\label{accHcombined}
\end{subequations}

The grouping of terms in these expressions can be understood as follows: relative to the leading dominant Newtonian acceleration, $GM/r^2$ for body $a$ and $Gm_b/r^2$ for the black hole, the BH terms are of order $GM/c^2r$ and involve the indirect effect of interations of stars with the black hole on its own motion, while the cross terms are of order $Gm_b /c^2r$ (recall that $v_b \sim (GM/r)^{1/2}$).  In addition the terms kept in the equation of motion for the black hole are necessary to ensure conservation of total momentum.   Although the terms neglected in each case are different in the context of the acceleration (see Eqs.\ (\ref{acc1}) and (\ref{accH})), in the context of the force on each star, $m_a {\bm a}_a$ and on the black hole $M{\bm a}_1$, the neglected terms are of the same order, $G^2 m_a^3/c^2 r^3$, in each case.  It is useful to point out that, in the case of two bodies, the terms involving summation over $b$ in Eqs.\ (\ref{acc1combined}) and those involving double summations in Eqs.\ (\ref{accHcombined}) drop out, and the remaining terms correspond to suitably truncated versions of the two-body equations of motion found, for example, in \cite{patiwill2,lucLRR}, with the replacement $\bm{v}_1 = -m_2 \bm{v}_2/M$.

It is straightforward, though tedious to show that these equations of motion can be derived from the truncated Lagrangian and Hamiltonian, given by
\begin{subequations}
\begin{eqnarray}
L &=& \frac{1}{2} \sum_a m_a v_a^2 + \frac{1}{2} Mv_1^2 + \frac{1}{2} \sum_{a,b} \frac{Gm_a m_b}{r_{ab}} + \sum_a \frac{GMm_a}{r_{1a}}
\nonumber \\
&& + \frac{1}{c^2} \biggl \{ \frac{1}{8} \sum_a m_a v_a^4 
+ \frac{3}{2} \sum_{a} \frac{G M m_a }{r_{1a}}  v_a^2
- \frac{1}{2} \sum_a \frac{G^2 M^2 m_a}{r_{1a}^2}
\nonumber \\
&& \quad
+ \frac{1}{2} \sum_{a,b} \frac{G m_a m_b}{r_{ab}} \left [ 3v_a^2 - 7 {\bm v}_a \cdot {\bm v}_b - ({\bm n}_{ab} \cdot {\bm v}_a )({\bm n}_{ab} \cdot {\bm v}_b ) \right ]
\nonumber \\
&& \quad
+ \frac{1}{2} \sum_{a} \frac{G M m_a }{r_{1a}} \left [ 3v_1^2 - 7 {\bm v}_1 \cdot {\bm v}_a
- ({\bm n}_{1a} \cdot {\bm v}_1 )({\bm n}_{1a} \cdot {\bm v}_a ) \right ]
\nonumber \\
&& \quad
- \sum_{a,b} \frac{G^2 Mm_a m_b }{r_{ab} r_{1a}}
- \frac{1}{2} \sum_{a,b} \frac{G^2 Mm_a m_b }{r_{1a} r_{1b}}
\biggr \} \,,
\label{NbodyL}
\\
H &=& \sum_a \frac{p_a^2}{2m_a} + \frac{p_1^2}{2M} 
-  \frac{1}{2} \sum_{a,b} \frac{Gm_a m_b}{r_{ab}} - \sum_a \frac{GMm_a}{r_{1a}}
\nonumber \\
&& - \frac{1}{c^2} \biggl \{ \frac{1}{8} \sum_a\frac{p_a^4}{m_a^3} + \frac{3}{2} \sum_a
\frac{GM}{r_{1a}} \frac{p_a^2}{m_a} 
- \frac{1}{2} \sum_a \frac{G^2 M^2 m_a}{r_{1a}^2}
\nonumber \\
&& \quad
+ \frac{1}{4} \sum_{a,b} \frac{G }{r_{ab}} \left [ 6\frac{m_b}{m_a} p_a^2 - 7 {\bm p}_a \cdot {\bm p}_b - ({\bm n}_{ab} \cdot {\bm p}_a )({\bm n}_{ab} \cdot {\bm p}_b ) \right ]
\nonumber \\
&& \quad
+ \frac{1}{2} \sum_{a} \frac{G  }{r_{1a}} \left [ 3\frac{m_a}{M} p_1^2 - 7 {\bm p}_1 \cdot {\bm p}_a
- ({\bm n}_{1a} \cdot {\bm p}_1 )({\bm n}_{1a} \cdot {\bm p}_a ) \right ]
\nonumber \\
&& \quad
- \sum_{a,b} \frac{G^2 Mm_a m_b }{r_{ab} r_{1a}}
- \frac{1}{2} \sum_{a,b} \frac{G^2 Mm_a m_b }{r_{1a} r_{1b}}
\biggr \} \,.
\label{NbodyH}
\end{eqnarray}
\end{subequations}
The first three PN terms in the Hamiltonian, Eq.\ (\ref{NbodyH}) correspond to the post-Newtonian Hamilton $H_{\rm PN}$ given in Eq.\ (27) of~\cite{1994AJ....108.1962S}.    However it is important to note that, because of the complex relationship between canonical momentum and velocity,
those terms yield not only the BH terms in the acceleration of body $a$, but also cross terms.   So a consistent Hamiltonian requires including the explicit cross terms which make up the remaining PN terms in Eq.\ (\ref{NbodyH}).

These equations admit the conserved total energy and momentum, given to the appropriate order by
\begin{subequations}
\begin{eqnarray}
E &=& \frac{1}{2} Mv_1^2 + \frac{1}{2} \sum_a m_a v_a^2 
-  \frac{1}{2} \sum_{a,b} \frac{Gm_a m_b}{r_{ab}} - \sum_a \frac{GMm_a}{r_{1a}}
\nonumber \\
&& + \frac{1}{c^2} \biggl \{ \frac{3}{8} \sum_a m_a v_a^4 + \frac{3}{2} \sum_a
\frac{GMm_a}{r_{1a}} v_a^2 
+ \frac{1}{2} \sum_a \frac{G^2 M^2 m_a}{r_{1a}^2}
\nonumber \\
&& \quad
+ \frac{1}{4} \sum_{a,b} \frac{G m_a m_b}{r_{ab}} \left [ 6v_{a}^2 - 7 {\bm v}_a \cdot {\bm v}_b - ({\bm n}_{ab} \cdot {\bm v}_a )({\bm n}_{ab} \cdot {\bm v}_b ) \right ]
\nonumber \\
&& \quad
+ \frac{1}{2} \sum_{a} \frac{G M m_a }{r_{1a}} \left [ 3v_1^2 - 7 {\bm v}_1 \cdot {\bm v}_a - ({\bm n}_{1a} \cdot {\bm v}_1 )({\bm n}_{1a} \cdot {\bm v}_a ) \right ]
\nonumber \\
&& \quad
+ \sum_{a,b} \frac{G^2 Mm_a m_b }{r_{ab} r_{1a}}
+ \frac{1}{2} \sum_{a,b} \frac{G^2 Mm_a m_b }{r_{1a} r_{1b}}
\biggr \} + O\left (\frac{G^2 m_a^3}{r^2} \right)\,, 
\\
{\bm P} &=& M{\bm v}_1 + \sum_{a} m_a \bm{v}_a \left ( 1 + \frac{1}{2c^2} v_a^2 \right )
- \frac{1}{2c^2} \biggl \{ \sum_a \frac{GMm_a}{r_{1a}} \left [ {\bm v}_a +  ({\bm v}_a \cdot {\bm n}_{1a})  {\bm n}_{1a} \right ]
\nonumber \\
&&
+ \sum_a \frac{GMm_a}{r_{1a}} \left [ {\bm v}_1 + ({\bm v}_1  \cdot {\bm n}_{1a})  {\bm n}_{1a} \right ]
+ \sum_{a,b} \frac{Gm_a m_b}{r_{ab}} \left [ {\bm v}_a +  ({\bm v}_a \cdot {\bm n}_{ab})  {\bm n}_{ab} \right ] \biggr \}
\nonumber \\
&& 
 + O\left (\frac{Gm_a^3 v_a}{c^2 Mr} \right ) \,,
\end{eqnarray}
\end{subequations}
where the sums exclude the black hole. 
These are the appropriately truncated versions of Eqs.\ (\ref{energymomentum}).
Note that, in the PN terms in $E$ and  $\bm P$, one could substitute the lowest-order relation ${\bm v}_1 =- \sum_b m_b {\bm v}_b /M$.   Finally the constant center of mass of the system can be defined to the same order:
\begin{eqnarray}
M^*\bm{X} &=& M{\bm x_1} \left ( 1 + \frac{1}{2c^2} v_1^2 \right )+  \sum_{a} m_a \bm{x}_a \left ( 1 + \frac{1}{2c^2} v_a^2 \right )
- \frac{1}{2c^2} \biggl \{ \sum_a \frac{GMm_a}{r_{1a}} \left ( {\bm x}_1 +  {\bm x}_a \right )
+ \sum_{a,b} \frac{Gm_a m_b}{r_{ab}} {\bm x}_a \biggr \}
\nonumber \\
&& 
 + O\left (\frac{Gm_a^3 x_a}{c^2 Mr} \right ) \,,
\end{eqnarray}
\end{widetext}
where $M^* = M + \sum_a m_a$ is the total mass of the system.  It is straightforward to show that $M^* d\bm{X}/dt = \bm{P}$.

With Kupi and Merritt, we have begun to devise possible numerical experiments to study the importance of these post-Newtonian cross terms in long-term evolutions of $N$-body systems around a central black hole.

\section{Post-Newtonian effects in hierarchical 3-body systems}
\label{sec:triples}

We now consider a three-body system in which two bodies of mass $m_1$ and $m_2$ are in a close orbit with separation $r$, and a third body of mass $m_3$ is in a wide orbit with separation $R \gg r$.   We define the relative separation vector of the two-body system and the vector from the center of mass of the two-body system to the third body by
\begin{equation}
{\bm x} \equiv {\bm x}_1 - {\bm x}_2 \,, \quad {\bm X} \equiv {\bm x}_3 - {\bm x}_0  \,,
\end{equation}
where
\begin{equation}
{\bm x}_0 \equiv \frac{m_1 {\bm x}_1 + m_2 {\bm x}_2}{m} \,,
\end{equation}
where $m \equiv m_1 + m_2$ is the mass of the two-body system.
We work in the center of mass-frame of the entire system, where 
\begin{equation}
m_1 {\bm x}_1 + m_2 {\bm x}_2 + m_3 {\bm x}_3 = m{\bm x}_0 + m_3 {\bm x}_3 = O(c^{-2}) \,,
\label{centerof mass}
\end{equation}
where $O(c^{-2})$ represents post-Newtonian corrections to the center of mass. 
As a result of these definitions, 
\begin{equation}
{\bm x}_1 =  \frac{m_2}{m} {\bm x} - \frac{m_3}{M} {\bm X} ,\, 
 {\bm x}_2 = - \frac{m_1}{m} {\bm x} - \frac{m_3}{M} {\bm X} ,\,
 {\bm x}_3 = \frac{m}{M} {\bm X} ,
\end{equation}
where $M = m_1 + m_2 + m_3$ is the total mass.  The  $O(c^{-2})$ correction in Eq.\ (\ref{centerof mass})  will not be relevant because only differences between position vectors appear in the equations of motion.  We define the velocities ${\bm v} \equiv d{\bm x}/dt$, ${\bm V} \equiv d{\bm X}/dt$, accelerations ${\bm a} \equiv d{\bm v}/dt$, ${\bm A} \equiv d{\bm V}/dt$, distances $r \equiv |{\bm x}|$, $R \equiv
|{\bm X}|$, and unit vectors ${\bm n} \equiv {\bm x}/r$ and ${\bm N} \equiv {\bm X}/R$. For future use we define the symmetric reduced mass $\eta \equiv m_1 m_2/m^2$ and the dimensionless mass difference $\Delta \equiv (m_1 - m_2)/m$.

We now turn to the EIH equations of motion (\ref{eq:EIH}), truncated to three bodies. 
At Newtonian order, the relative acceleration within the two-body system has terms of order
$Gm/r^2$ and $Gm_3 r/R^3$, where we have expanded the effect of the  external body to only quadrupole order (in principle the expansion can be carried out to higher orders).
In the post-Newtonian approximation, each of these terms would be represented by a relativistic potential $Gm/rc^2$ and $Gm_3 r^2/R^3c^2$.  Thus we wish to keep the standard Newtonian terms, plus PN correction terms from the two-body system, of order $G^2m^2/r^3 c^2$, {\em plus} PN--third-body cross terms of the form $(Gm/r^2)\times(Gm_3r^2/R^3c^2)$ or $(Gm_3r/R^3)\times(Gm/rc^2)$, both of which scale as $G^2 mm_3/R^3c^2$.
Because the PN equations also contain velocity-dependent terms, with $v \sim (Gm/r)^{1/2}$ and $V \sim (Gm/R)^{1/2}$, there will also be cross terms that scale as 
$(G^2 mm_3/R^3c^2)(R/r)^{n/2}$, where $n$ ranges from $1$ to $5$.  All cross terms that scale as $(G^2 mm_3 r/R^4c^2)$ or higher powers of $r$ will be dropped.

For a consistent set of equations of motion that maintain the globally conserved energy and linear momentum to the appropriate order, we must also include the equations of motion for body $3$.   Since momentum conservation requires that $m {\bm a} \sim m_3 {\bm a}_3$, we must therefore include in those equations of motion terms that scale as $m/m_3$ compared to those in the two-body equations, in other words, terms that scale as $(G^2 m^2/R^3 c^2)(R/r)^{n/2}$, where $n$ will turn out to range from $0$ to $4$.

The interaction of the two bodies with the third body depends on ${\bm x}_{13}$ and ${\bm x}_{23}$, which we will express as
\begin{eqnarray}
{\bm x}_{13} &=&  -{\bm X} + \alpha_2 {\bm x} = -R \left [{\bm N} - \alpha_2 (r/R) {\bm n}\right ]\,,
\nonumber \\
{\bm x}_{23} &=&  -{\bm X} - \alpha_1 {\bm x} = -R \left [{\bm N} + \alpha_1 (r/R) {\bm n}\right ]\,, 
\end{eqnarray}
where $\alpha_i \equiv m_i/m$; we will use this to expand quantities such as $1/r_{13}$ and $1/r_{23}$ as power series in $r/R$.  The resulting equations of motion for the binary system have the form, 
\begin{eqnarray}
{\bm a} &=& - \frac{Gm{\bm n}}{r^2} - \frac{Gm_3 \,r}{R^3} \left [ {\bm n} -3({\bm n} \cdot {\bm N}) {\bm N} \right ]
+ \frac{1}{c^2}[{\bm a}]_{\rm Binary} 
\nonumber \\
&& \quad + \frac{1}{c^2}[{\bm a}]_{\rm Cross}
+ O\left (\frac{G^2 m m_3 r}{c^2 R^4}\right ) \,,
\label{acc13}
\end{eqnarray}
where we have expanded the Newtonian term from the third body to quadrupole order, and where the Binary and Cross terms are given by
\begin{widetext}
\begin{subequations}
\begin{eqnarray}
[{\bm a}]_{\rm Binary} &=& \frac{Gm{\bm n}}{r^2} \left [(4+2\eta) \frac{Gm}{r} - (1+3\eta)v^2 + \frac{3}{2} \eta \dot{r}^2 \right ] + (4-2\eta) \frac{Gm\dot{r} {\bm v}}{r^2} \,,
 \label{eqn:abinary}
 \\
\left [{\bm a}\right]_{\rm Cross} &=& \frac{Gm_3 \Delta}{r^2} \left [ 2{\bm n} ({\bm v} \cdot {\bm V}) + {\bm v}  ({\bm n} \cdot {\bm V}) \right ]
 + 5 \frac{G^2m m_3 {\bm n}}{r^2 R}
 \nonumber \\
&&
+ \frac{Gm_3 \Delta}{R^2} \left [ \frac{1}{2}\frac{Gm}{r} \left \{ {\bm N} - 9({\bm n} \cdot {\bm N}) {\bm n} \right \} + 4 {\bm v} ({\bm N} \cdot {\bm v}) - v^2 {\bm N} \right ]
 \nonumber \\
&&
-\frac{Gm_3}{R^2} \left [ 4{\bm v} \times ({\bm N} \times {\bm V}) -3{\bm v} ({\bm N} \cdot {\bm V})  \right ]
 \nonumber \\
&&
+\frac{G^2 mm_3}{R^3}\left [ (4-\eta)\left \{ {\bm n} -3({\bm n} \cdot {\bm N}) {\bm N} \right \}
-\frac{1}{2} (4 -13\eta) {\bm n} \left \{ 1- 3({\bm n} \cdot {\bm N})^2 \right \} \right ]
 \nonumber \\
&&
+ \frac{Gm_3 r}{R^3} (1-3\eta) \left [ 4{\bf v} \left \{ \dot{r} -3({\bm n} \cdot {\bm N})({\bm v} \cdot {\bm N}\right \} - v^2 \left \{ {\bm n} -3({\bm n} \cdot {\bm N}) {\bm N} \right \} \right ]
  \,,
  \label{eqn:across}
\end{eqnarray}
\label{aPN3body}
\end{subequations}
where $\dot{r} \equiv {\bm n} \cdot {\bm v}$.
Recalling that $v^2 \sim Gm/r$, and $V^2 \sim Gm/R$ we see that the six cross terms scale as $(Gmm_3/R^3c^2) \times (R/r)^{n/2}$, where $n = 5,\, 4,\, 2,\, 1,\, 0,\, 0$, respectively.

Treating the third body in the analogous way and defining ${\bm A} \equiv d^2 {\bm X}/dt^2$, we obtain
\begin{equation}
{\bm A} = - \frac{GM{\bm N}}{R^2} + \frac{3}{2}\frac{GM \eta r^2}{R^4} \left [ {\bm N} \left (1- 5({\bm n} \cdot {\bm N})^2 \right) + 2{\bm n} ({\bm n} \cdot {\bm N})\right ]
+ \frac{1}{c^2}[{\bm A}]_{\rm Binary} + \frac{1}{c^2}[{\bm A}]_{\rm Cross}
+ O\left (\frac{G^2 m^2 r}{c^2 R^4}\right ) \,,
\label{acc3}
\end{equation}
where
\begin{eqnarray}
[{\bm A}]_{\rm Binary} &=& \frac{Gm{\bm N}}{R^2} \left [ \frac{4Gm}{R} - V^2  \right ] + \frac{4Gm({\bm N} \cdot {\bm V}){\bm V}}{R^2} \,,
 \\
\left [{\bm A}\right]_{\rm Cross} &=&
- \frac{Gm \eta \Delta}{r^2} \left [ {\bm n} \left ( \frac{Gm}{r} -v^2 + \frac{3}{2} \dot{r}^2 \right ) - \dot{r} {\bm v} \right ]
\nonumber \\
&& 
+\frac{1}{2}\frac{G^2 m^2 \eta}{rR^2} \left [ {\bm N} \left (5-3({\bm n} \cdot {\bm N})^2 \right ) - 6{\bm n}({\bm n} \cdot {\bm N}) \right ]
\nonumber \\
&&
- \frac{Gm\eta}{R^2} \left [ 2v^2 {\bm N} - \frac{3}{2} {\bm N}({\bm v} \cdot {\bm N})^2
   - 3 {\bm v} ({\bm v} \cdot {\bm N}) \right ]
\nonumber \\
&&
- \frac{1}{4} \frac{G^2 m^2 \eta \Delta}{R^3} \left [ {\bm n} \left (1-15 ({\bm n} \cdot {\bm N})^2 \right ) + 3 {\bm N}({\bm n} \cdot {\bm N}) \left ( 7 -5({\bm n} \cdot {\bm N})^2 \right ) \right ]  
\nonumber \\
&&
- \frac{Gm\eta \Delta \, r}{R^3} \biggl [ 2v^2 \left ( {\bm n} -3{\bm N}({\bm n} \cdot {\bm N}) \right ) - \frac{3}{2} \left ( {\bm n} ({\bm v} \cdot {\bm N})^2 + 2 {\bm N} \dot{r} ({\bm v} \cdot {\bm N})
-5 {\bm N} ({\bm n} \cdot {\bm N}) ({\bm v} \cdot {\bm N})^2  \right ) 
\nonumber \\
&&  \qquad - 3{\bm v} \left (\dot{r} - 3({\bm v} \cdot {\bm N}) ({\bm n} \cdot {\bm N}) \right ) \biggr ] \,.
\end{eqnarray}
\end{widetext}
The Binary--PN acceleration contains the standard terms for a body in orbit around a point mass of mass $m$, but in the approximations used here it has the same scaling as the cross terms shown in $[{\bm A}]_{\rm Cross}$.  The first term in $[{\bm A}]_{\rm Cross}$ comes from the acceleration of the center of mass ${\bm x}_0$ of the binary.

In order to illustrate the impact of the cross terms on the dynamics, we 
apply these equations to a specific simple case.  We consider the third body to be in a circular orbit on the $X-Y$ plane, with $R$ constant, angular frequency $\Omega_3 = (Gm/R^3)^{1/2}$ and with $\bm{X} = R \bm{N}$ and $\bm{V} = \Omega_3 R \bm{\Lambda}$, where 

\begin{eqnarray}
\bm{N} &=& \bm{e}_X \cos \Omega_3 t + \bm{e}_Y \sin \Omega_3 t \,,
\nonumber \\
\bm{\Lambda} &=& -\bm{e}_X \sin \Omega_3 t + \bm{e}_Y \cos \Omega_3 t \,,
\nonumber \\
\bm{H} &=& \bm{N} \times \bm{\Lambda} = \bm{e}_Z \,.
\end{eqnarray}

The relative osculating orbit of the two-body system is governed by Eqs.\ (\ref{keplerorbit}), with osculating orbit elements $a$, $e$, $\omega$, $\Omega$ and $\iota$, with $a \ll R$.  We focus on perturbations of the binary orbit induced by the third body.  The
radial $\cal R$, cross-track $\cal S$ and out-of-plane $\cal W$ components of the perturbing acceleration are given by
\begin{widetext}
\begin{subequations}
\begin{eqnarray}
{\cal R} &=& - \frac{Gm_3 r}{R^3} \left (1-3(\bm{n} \cdot \bm{N})^2 \right ) + \frac{Gm}{r^2 c^2} \left [ 2(2+\eta) \frac{Gm}{r} - (1+3\eta)v^2 + \frac{1}{2}(8-\eta) \dot{r}^2 \right ]
\\
\nonumber &&
+ \frac{Gm_3}{c^2} \biggl \{ \frac{ \Delta}{r^2} (2\bm{v} \cdot \bm{V}+ \dot{r} \bm{n} \cdot \bm{V})+ 5 \frac{Gm}{r^2 R}
- \frac{\Delta}{R^2} \left [\left ( 4\frac{Gm}{r}+ v^2 \right ) \bm{n}\cdot \bm{N} - 4\dot{r} \bm{v} \cdot \bm{N}  \right ]
\\
\nonumber &&
- \frac{4}{R^2} (\bm{n} \times \bm{v}) \cdot (\bm{N} \times \bm{V})
+ 4 (1-3\eta) \frac{r\dot{r}}{R^3} \left (\dot{r} - 3(\bm{n} \cdot \bm{N})(\bm{v} \cdot \bm{N}) \right )
\\
\nonumber &&
+ \frac{r}{2R^3} \left [ (4+11\eta)\frac{Gm}{r} -  2(1-3\eta)v^2 \right ] \left ( 1-3(\bm{n} \cdot \bm{N})^2 \right )
\biggr \} \,,
\\
{\cal S} &=& 3\frac{Gm_3 r}{R^3} (\bm{n} \cdot \bm{N})(\bm{\lambda} \cdot \bm{N})
+ 2(2-\eta) \frac{Gm}{c^2 r^2} h \dot{r}
\\
\nonumber &&
+ \frac{Gm_3}{c^2} \biggl \{ 
\frac{h \Delta}{r^3}  \bm{n} \cdot \bm{V}
+ \frac{\Delta}{R^2} \left [ \left ( \frac{Gm}{2r} -v^2 \right ) \bm{\lambda} \cdot \bm{N} +  \frac{4h}{r} \bm{v} \cdot \bm{N} \right ]
\\
\nonumber &&
- \frac{4}{R^3} (\bm{\lambda} \times \bm{v}) \cdot (\bm{N} \times \bm{V})
+ 4(1-3\eta) \frac{h}{R^3}  \left (\dot{r} - 3(\bm{n} \cdot \bm{N})(\bm{v} \cdot \bm{N}) \right )
\\
\nonumber &&
- \frac{3r}{R^3} \left [ (4-\eta)\frac{Gm}{r} - (1-3\eta)v^2 \right ] (\bm{n} \cdot \bm{N})(\bm{\lambda} \cdot \bm{N})
\biggr \} \,,
\\
{\cal W} &=& 3\frac{Gm_3 r}{R^3} (\bm{n} \cdot \bm{N})(\bm{\hat{h}} \cdot \bm{N})
\\
\nonumber &&
+ \frac{Gm_3}{c^2} \biggl \{ 
+  \frac{\Delta}{R^2} \left ( \frac{Gm}{2r} -v^2 \right ) \bm{\hat{h}} \cdot \bm{N}
- \frac{4}{R^3} (\bm{\hat{h}} \times \bm{v}) \cdot (\bm{N} \times \bm{V})
\\
\nonumber &&
- \frac{3r}{R^3} \left [   (4-\eta)\frac{Gm}{r} -(1-3\eta) v^2 \right ](\bm{n} \cdot \bm{N})(\bm{\hat{h}} \cdot \bm{N})
\biggr \} \,,
\end{eqnarray}
\label{kozaiRSW}
\end{subequations}
where we recall that $\bm{h} \equiv \bm{x} \times \bm{v}$, and $\bm{v} = \dot{r} \bm{n} + (h/r) \bm{\lambda}$.

Because the orbital period of the inner binary is assumed to be short compared to that of the third body, it is customary to integrate the Lagrange planetary equations holding the position of the third body fixed, and then to average over one orbit of the third body.  This yields a set of averaged equations for the rates of change of the orbit elements that can then be evolved over longer timescales.  

Looking first at the Newtonian and post-Newtonian binary terms we recover the standard results for the Kozai problem, including the two-body pericenter advance:
\begin{subequations}
\begin{eqnarray}
\langle \Delta a \rangle_{\rm K} &=& 0 \,,
\label{delakozai}
\\
\langle \Delta e  \rangle_{\rm K}&=& \frac{15\pi}{2} \frac{m_3}{m} \left ( \frac{a}{R} \right )^3 e(1-e^2)^{1/2} \sin^2 \iota \sin \omega \, \cos \omega \,,
\label{delekozai}
\\
\langle \Delta \iota \rangle_{\rm K} &=& -\frac{15\pi}{2} \frac{m_3}{m} 
\left( \frac{a}{R} \right)^3 e^2 (1-e^2)^{-1/2} \sin\iota \cos\iota  
\sin\omega \cos\omega \,,
\label{deliotakozai}
\\
\sin \iota \langle \Delta \Omega \rangle_{\rm K} &=& -\frac{3\pi}{2} \frac{m_3}{m} 
\left( \frac{a}{R} \right)^3 e^2 (1-e^2)^{-1/2}(1-e^2 + 5e^2 \sin^2 \omega ) \sin\iota \cos\iota   \,,
\label{delOmegakozai}
\\ 
\langle \Delta \omega \rangle_{\rm K} &=&  \frac{6\pi Gm}{c^2a(1-e^2)}
+\frac{3\pi}{2} \frac{m_3}{m}  
\left ( \frac{a}{R} \right)^3 (1-e^2)^{-1/2} \left [ 5 \cos^2\iota 
\sin^2\omega + (1-e^2)(5\cos^2\omega - 3) \right ]  \,.  
\label{delomegakozai}
\end{eqnarray}
\label{delKozai}
\end{subequations}

We determine the effect of the cross terms using the method outlined in Sec.\ \ref{sec:conundrum}.  Taking the Newtonian and PN binary contributions to the disturbing accelerations, we calculate the instantaneous values of the osculating elements as a function of $f$, again holding the position of the third body fixed.  We insert these expressions back into the formulae for $r$, $\bm{v}$, $\bm{n}$, $\bm{h}$ that appear in Eqs.\ (\ref{kozaiRSW}), and keep contributions that involve cross terms between $m_3$ and $1/c^2$  We also insert the Newtonian and PN binary expressions for 
$\dot{\omega}$ and $\dot{\Omega}$ into the formula (\ref{dfdt}), used to convert derivatives with respect to time in the Lagrange planetary equations to derivatives with respect to true anomaly $f$, again keeping the appropriate cross terms.  These are combined with the explicit cross terms that appear in Eqs.\ (\ref{kozaiRSW}).   We then integrate over a complete binary orbit and average over the position of the third body.  The resulting average changes in the orbit elements are given by
\begin{subequations}
\begin{eqnarray}
\langle \Delta a \rangle_{\rm Cross} &=& -\frac{15\pi}{2}\frac{Gm_3}{c^2} \left ( \frac{a}{R} \right )^3  \biggl \{ \frac{e(1+e)^2}{(1-e)(1-e^2)^{3/2}} \left [ 7+3e-\eta(3+4e) \right ]  +\frac{6}{5}\frac{1-e}{1+e} \biggr \} \sin^2 \iota \sin 2\omega \,,
\label{delacross}
\\
\langle \Delta e \rangle_{\rm Cross} &=& - \frac{15\pi}{8}\frac{Gm_3}{a c^2} \left ( \frac{a}{R} \right )^3 \biggl \{ \biggl [ \frac{(1+e)^2}{(1-e)(1-e^2)^{1/2}} \left \{ (3+7e)-(1+6e)\eta -f(e,\eta) \right \} 
\nonumber \\
&& \qquad \qquad
+\frac{4}{5} \frac{(1-e)^2(2+4e-3e^2)}{e^3} \biggr ] \sin 2\omega  
- 12\pi \frac{e}{(1-e^2)^{1/2}}\cos 2 \omega  \biggr \} \sin^2 \iota  \,,
\label{delecross}
\\
\langle \Delta \iota \rangle_{\rm Cross} &=&- \frac{15\pi}{8}\frac{Gm_3}{a c^2} \left ( \frac{a}{R} \right )^3 \biggl \{ \biggl [ \frac{e(1+e)^2}{(1-e)(1-e^2)^{3/2}} \left \{ (3+7e)-(1+6e)\eta + f(e,\eta) \right \} 
\nonumber \\
&& \qquad \qquad
-\frac{8}{5} \frac{(1-e)^3(1+3e)}{e^2(1-e^2)} \biggr ] \sin 2\omega 
+ 12\pi \frac{e^2}{(1-e^2)^{3/2}}\cos 2 \omega  \biggr \} \sin \iota \cos \iota \,,
\label{deliotacross}
\\
\sin \iota \langle \Delta \Omega \rangle_{\rm Cross} &=&  
4\pi \frac{Gm_3}{a c^2} \left ( \frac{a}{R} \right )^{5/2} \sin \iota  
- \frac{3\pi}{8}\frac{Gm_3}{a c^2} \left ( \frac{a}{R} \right )^3  \frac{(1+e)}{(1-e)(1-e^2)^{3/2}} g(e,\eta) \sin \iota \cos \iota
\nonumber \\
&& \qquad \qquad
+  \langle \Delta \iota \rangle_{\rm Cross} (\omega \to \omega - \pi/4 ) 
 \,,
 \label{delOmegacross}
\end{eqnarray}
\label{delcross}
\end{subequations}
where
\begin{eqnarray}
f(e,\eta) &\equiv& \frac{8-16e-24e^2+109e^3+114e^4+43e^5+16e^6-\eta e^3(15+47e+76e^2+37e^3)}{5e^3(1+e)} \,,
\nonumber \\
g(e,\eta) &\equiv&  2(2+3e)(12+12e+11e^2)-\eta(24+40e+84e^2+86e^3+11e^4) \,.
\end{eqnarray}
\end{widetext}

The first three terms in $\bm{a}_{\rm Cross}$, corresponding to $n = 5,\, 4$ and $2$, do not generate a secular change in any orbital element.  The fourth term, corresponding to $n =1$, generates the first term
in $\sin \iota \langle \Delta \Omega \rangle_{\rm Cross}$.  This term is the analogue of the de Sitter precession, a relativistic contribution to the nodal precession in the Earth-Moon system induced by the Sun, that has been measured to a few percent using lunar laser ranging.  Only the final $n=0$ terms in $\bm{a}_{\rm Cross}$ generate other secular changes in orbit elements.

Notice that, unlike the standard Kozai case, where the semi-major axis is constant over an orbit, including the cross terms leads to a variation in $a$ proportional to $\sin^2 \iota$ and to $\sin 2\omega$.   

In the standard Kozai problem, $L_Z \equiv (\bm{x} \times \bm{v} ) \cdot \bm{e}_Z = [Gma(1-e^2)]^{1/2} \cos \iota$ is constant, as can be seen from Eqs.\ (\ref{delakozai}), (\ref{delekozai}) and (\ref{deliotakozai}).  But the cross terms now induce a change in $L_Z$ over one orbit given, from Eqs. (\ref{delacross}), (\ref{delecross}) and (\ref{deliotacross}) by
\begin{eqnarray}
 \frac{\langle \Delta L_Z \rangle_{\rm Cross}}{L_Z} &=& -\frac{15\pi}{2} \frac{Gm_3}{a c^2} \left ( \frac{a}{R} \right )^{3} \frac{e(1+e)^2}{(1-e^2)^{3/2}} 
 \nonumber \\
 && \quad \times(2-\eta) \sin^2 \iota \sin 2\omega \,.
 \label{Lzchange}
\end{eqnarray}
Since we are averaging over the circular orbit of the third body, we would expect the $z$-component of the angular momentum of the two-body system still to be conserved, because of the axial symmetry of the ``averaged'' perturbation.  However the quantity that is conserved contains post-Newtonian corrections, which must be taken into account.  Starting with $\bm{x} \times d\bm{v}/dt$ from the equations of motion (\ref{acc13}) and (\ref{aPN3body}) and contracting it with $\bm{H}=\bm{e}_Z$, it is straightforward to show, after averaging over the orbit of the third body, that the following quantity is a constant of the motion:
\begin{eqnarray}
\tilde{L}_Z &=& L_Z  \left ( 1+ 2\frac{Gm}{rc^2} (2-\eta) \right )
\nonumber \\
&&
- 2 L_Z \frac{Gm_3 r^2}{R^3 c^2}(1-3\eta) \left [ 1-3(\bm{n} \cdot \bm{N})^2 \right ]
\nonumber \\
&& - 2 \frac{Gm_3 r^2 \Delta}{R^3 c^2} \left [1 - (\bm{n} \cdot \bm{e}_Z)^2 \right ]  \,.
 \label{eq:Lzconserved}
\end{eqnarray}
To the order of approximation considered, the second and third terms of Eq. (\ref{eq:Lzconserved}) are periodic or constant, so their change over one orbit of the binary is zero.  However, the term $Gm/rc^2$ in the first term is not strictly periodic.  Because the orbit element  $e$ changes over one orbit by an amount of order $(m_3/m)(a/R)^3$ from the standard Kozai mechanism, it induces a net change in $1/r$ over an orbit.  Thus, holding $\tilde{L}_Z$ fixed, and determining the variation between successive pericenters we find that 
\begin{eqnarray}
 \frac{\langle \Delta L_Z \rangle_{\rm Cross}}{L_Z} &\approx& -2 \frac{Gm}{c^2} (2-\eta) \Delta \left ( \frac{1}{r} \right )
 \nonumber \\
 && = -2 \frac{Gm}{a c^2} (2-\eta) \Delta \left ( \frac{1}{1-e} \right )
  \nonumber \\
 && = -2 \frac{Gm}{a c^2} (2-\eta) \frac{\Delta e}{(1-e)^2} \,,
\end{eqnarray}
Substituting for $\Delta e$ from Eq. (\ref{delekozai}) yields Eq. (\ref{Lzchange}).  

Naoz et al.~\cite{2013ApJ...773..187N} used a complete three-body PN Hamilton converted to Delaunay variables, followed by a double orbit average to obtain evolution equations for the variables that include cross-term effects of the same scaling as those above.   However a comparison between the two methods is not trivial, as the Delaunay variables and the osculating orbit elements are very different.  For example, our $L_Z$ is defined relative to the reference $XYZ$ system, defined so that the third body is initially on the $XY$ plane, whereas the analogous Delaunay variable $H_1$ is the component along the total angular momentum vector of the system.    On the other hand, it is not clear whether the double orbit average in the Delaunay method adequately takes into account the feedback of periodic terms in the first-order solutions for the Delaunay variables into the cross-term effects, as discussed in points 1 and 2 in Sec. \ref{sec:conundrum}.  This will be the subject of future investigations.

If the relativistic pericenter precession (the first term in Eq.\ (\ref{delomegakozai})) dominates the Kozai precession (the second term), then over a pericenter precession timescale, we can integrate Eq.\ (\ref{Lzchange}) using Eq.\ (\ref{integrate_element}) to obtain
\begin{eqnarray}
L_Z &=& (L_Z)_0 \biggl [ 1 - \frac{5}{4} \frac{m_3}{m} \left ( \frac{a}{R} \right )^3 
\frac{e(1+e)^2}{(1-e^2)^{1/2}}
\nonumber \\
\qquad &&  \times (2-\eta) \sin^2 \iota \sin^2\omega \biggr ] \,.
\label{Lzfinal}
\end{eqnarray}
Just as in the quadrupole case of Sec.\ \ref{sec:conundrum}, we find that post-Newtonian cross-terms induce a {\em Newtonian}-order variation in an orbital quantity --- here the angular momentum --- over a relativistic precession timescale.  The amplitude is of the same order as the {\em per orbit} variations in $e$, $\iota$ and $\Omega$ induced by the Newtonian Kozai mechanism.  Whether these variations have important long-term consequences for the complex three-body dynamics of Kozai systems can probably only be answered by numerical experiments using Eqs.\ (\ref{delKozai}) and (\ref{delcross}).

\section{Concluding remarks}
\label{sec:conclusions}

In this paper we have suggested that a proper approach to including post-Newtonian relativistic effects in the dynamics of $N$-body  systems that include either a central massive black hole or a close binary star system requires incorporating certain PN ``cross terms'' in the equations of motion.   There are two sides to this issue.

The first has to do with consistency.  The underlying problem is a mathematical one that is broader than $N$-body gravitational dynamics.   Here one begins with a simple system with an exact solution --- the Newtonian two-body system of star and black hole, or the inner binary system of a hierarchical triple  --- that is perturbed by two separate mechanisms.   A simple mathematical model that illustrates this is the differential equation $dx /dt = {\rm i} x + \epsilon x^2 + \delta x^3$, where $\epsilon$ and $\delta$ are small parameters.  In the gravitational problems discussed here, $\epsilon \sim 1/c^2$ characterizes the relativistic perturbations, either due to the central black hole or within the close binary system, while $\delta \sim m_b$ characterizes the Newtonian perturbation from the other body or bodies, or the Newtonian quadrupole perturbation in the case of our quadrupole conundrum.    In the absence of the perturbation, the solution of the differential equation with unit amplitude is $x = e^{{\rm i}t}$.   An ``exact'' solution of the full equation can be provided by a numerical integration.  But it is clear that, if one seeks a perturbative solution by trying an expansion of the form
$x(t) = e^{{\rm i}t} + \epsilon x_1 + \delta x_2 + \dots$, the series will necessarily have terms proportional to the product $\epsilon \delta$.  This is because changes in $x$ induced by the $\delta$-perturbing term, will induce a change in the $\epsilon$-perturbing term that will lead to a perturbation of order $\epsilon \delta$, and vice versa.  
We saw this phenomenon in action already in Sec.\ \ref{sec:conundrum}, where relativistic effects on the orbit elements induced changes in the quadrupole perturbing terms and vice versa.  We saw it again in Sec.\ \ref{sec:triples}.  Now, for the simple differential equation displayed above, a numerical integration will capture all such effects automatically.  However, if the equation to be solved also has {\em explicit} terms of order $\epsilon \delta$, then a failure to include them in the numerical integration will lead to a solution that is incomplete, and thus potentially incorrect.   We found exactly this phenomenon in the quadrupole conundrum.  A numerical integration of the equations that included only the ``monopole'' PN terms ($\epsilon$ terms) and the Newtonian quadrupole terms ($\delta$ terms) in the equations of motion did give a variation in the semi-major axis, via the mechanism described above.  However the answer was wrong.  {\em Only} when the explicit PN-quadrupole $\epsilon \delta$ cross terms in the equations of motion were included in the numerical integration was the correct answer obtained. 

Conversely, if one is handed an equation with explicit cross terms, such as $dx /dt = {\rm i} x + \epsilon x^2 + \delta x^3 + \epsilon \delta f(x)$, and attempts a perturbative solution, it is essential to plug the first order solution $x = e^{{\rm i}t} + \epsilon x_1 + \delta x_2$ back into the equation, thereby generating the additional $\epsilon \delta$ contributions.

For problems involving $N$ bodies orbiting a massive black hole, we have accordingly written down the required cross terms to an appropriate order, so as to guarantee consistency in the numerical solutions at the level of $\epsilon \delta$ effects.  
For hierarchical triple systems we have written down the analogous equations and have also carried out the orbital perturbation theory analytically for a simple case of a circular third-body orbit, yielding doubly orbit-averaged equations for the orbit elements of the inner binary, including the $\epsilon \delta$ effects in a completely consistent way.   

For triple systems, these $\epsilon \delta$ issues can be accounted for automatically by simply integrating the full EIH three-body equations of motion numerically.   In other $N$-body situations, where $N$ is sufficiently large, this may not be computationally practical, and thus our truncated equations may prove to be useful.

The second side to this issue is physical relevance:  do these ``cross terms'' produce physically interesting effects?  In principle they are smaller (of order $\epsilon \delta$) than either the relativistic ($\epsilon$) or $N$-body ($\delta$) effects, although we have seen that in some cases, over a relativistic precession timescale, they can be ``boosted'' to a size of order $\delta$.  The only way to answer this question is to carry out suitable numerical experiments.   Given the complexity of all gravitating $N$-body systems when $N > 2$, and given that seemingly tiny effects can over time yield dramatic consequences for the evolution of large systems, the cross terms emphasized in this paper could prove to have interesting consequences.

\acknowledgments

This work was supported in part by the National Science Foundation,
Grant Nos.\  PHY 09--65133, 12--60995 \& 13--06069.   We are grateful for the hospitality of  the Institut d'Astrophysique de Paris and  the Weizmann Institute of Science, Israel, where parts of this work were carried out.   We also thank the organizers of 
the ``Al\'ajar Meeting 2013: Stellar dynamics and growth of massive black holes''  for a stimulating venue for discussions of issues related to this work.   We thank David Merritt and Tal Alexander for patiently introducing us to key concepts and problems in $N$-body dynamics, Smadar Naoz for useful discussions of the Delaunay approach to hierarchical triple systems, and Lorenzo Iorio for directing our attention to Refs.~\cite{1988CeMec..42...81S,1991ercm.book.....B}. 

\appendix

\bibliography{refs}

\begin{thebibliography}{30}
\expandafter\ifx\csname natexlab\endcsname\relax\def\natexlab#1{#1}\fi
\expandafter\ifx\csname bibnamefont\endcsname\relax
  \def\bibnamefont#1{#1}\fi
\expandafter\ifx\csname bibfnamefont\endcsname\relax
  \def\bibfnamefont#1{#1}\fi
\expandafter\ifx\csname citenamefont\endcsname\relax
  \def\citenamefont#1{#1}\fi
\expandafter\ifx\csname url\endcsname\relax
  \def\url#1{\texttt{#1}}\fi
\expandafter\ifx\csname urlprefix\endcsname\relax\def\urlprefix{URL }\fi
\providecommand{\bibinfo}[2]{#2}
\providecommand{\eprint}[2][]{\url{#2}}

\bibitem[{\citenamefont{{Einstein}}(1939)}]{Einstein39}
\bibinfo{author}{\bibfnamefont{A.}~\bibnamefont{{Einstein}}},
  \bibinfo{journal}{Ann. Math.} \textbf{\bibinfo{volume}{40}},
  \bibinfo{pages}{922} (\bibinfo{year}{1939}).

\bibitem[{\citenamefont{Hoyle and Fowler}(1967)}]{HoyleFowler67}
\bibinfo{author}{\bibfnamefont{F.}~\bibnamefont{Hoyle}} \bibnamefont{and}
  \bibinfo{author}{\bibfnamefont{W.~A.} \bibnamefont{Fowler}},
  \bibinfo{journal}{Nature} \textbf{\bibinfo{volume}{213}},
  \bibinfo{pages}{373} (\bibinfo{year}{1967}).

\bibitem[{\citenamefont{{Ipser}}(1969)}]{1969ApJ...158...17I}
\bibinfo{author}{\bibfnamefont{J.~R.} \bibnamefont{{Ipser}}},
  \bibinfo{journal}{\apj} \textbf{\bibinfo{volume}{158}}, \bibinfo{pages}{17}
  (\bibinfo{year}{1969}).

\bibitem[{\citenamefont{{Alexander}}(2005)}]{2005PhR...419...65A}
\bibinfo{author}{\bibfnamefont{T.}~\bibnamefont{{Alexander}}},
  \bibinfo{journal}{\physrep} \textbf{\bibinfo{volume}{419}},
  \bibinfo{pages}{65} (\bibinfo{year}{2005}), \eprint{astro-ph/0508106}.

\bibitem[{\citenamefont{{Genzel} et~al.}(2010)\citenamefont{{Genzel},
  {Eisenhauer}, and {Gillessen}}}]{2010RvMP...82.3121G}
\bibinfo{author}{\bibfnamefont{R.}~\bibnamefont{{Genzel}}},
  \bibinfo{author}{\bibfnamefont{F.}~\bibnamefont{{Eisenhauer}}},
  \bibnamefont{and}
  \bibinfo{author}{\bibfnamefont{S.}~\bibnamefont{{Gillessen}}},
  \bibinfo{journal}{Rev. Mod. Phys.} \textbf{\bibinfo{volume}{82}},
  \bibinfo{pages}{3121} (\bibinfo{year}{2010}), \eprint{1006.0064}.

\bibitem[{\citenamefont{{Merritt}}(2013)}]{2013degn.book.....M}
\bibinfo{author}{\bibfnamefont{D.}~\bibnamefont{{Merritt}}},
  \emph{\bibinfo{title}{{Dynamics and Evolution of Galactic Nuclei}}}
  (\bibinfo{publisher}{Princeton University Press},
  \bibinfo{address}{Princeton}, \bibinfo{year}{2013}).

\bibitem[{\citenamefont{{Gezari}}(2013)}]{2013BrJPh.tmp...36G}
\bibinfo{author}{\bibfnamefont{S.}~\bibnamefont{{Gezari}}},
  \bibinfo{journal}{Brazilian Journal of Physics: Online First}
  (\bibinfo{year}{2013}).

\bibitem[{\citenamefont{{Amaro-Seoane}}(2012)}]{2012arXiv1205.5240A}
\bibinfo{author}{\bibfnamefont{P.}~\bibnamefont{{Amaro-Seoane}}},
  \bibinfo{journal}{ArXiv e-prints}  (\bibinfo{year}{2012}),
  \eprint{1205.5240}.

\bibitem[{\citenamefont{{Will}}(2008)}]{2008ApJ...674L..25W}
\bibinfo{author}{\bibfnamefont{C.~M.} \bibnamefont{{Will}}},
  \bibinfo{journal}{\apjl} \textbf{\bibinfo{volume}{674}}, \bibinfo{pages}{L25}
  (\bibinfo{year}{2008}), \eprint{0711.1677}.

\bibitem[{\citenamefont{{Rauch} and {Tremaine}}(1996)}]{1996NewA....1..149R}
\bibinfo{author}{\bibfnamefont{K.~P.} \bibnamefont{{Rauch}}} \bibnamefont{and}
  \bibinfo{author}{\bibfnamefont{S.}~\bibnamefont{{Tremaine}}},
  \bibinfo{journal}{\na} \textbf{\bibinfo{volume}{1}}, \bibinfo{pages}{149}
  (\bibinfo{year}{1996}), \eprint{astro-ph/9603018}.

\bibitem[{\citenamefont{{Hopman} and {Alexander}}(2006)}]{2006ApJ...645.1152H}
\bibinfo{author}{\bibfnamefont{C.}~\bibnamefont{{Hopman}}} \bibnamefont{and}
  \bibinfo{author}{\bibfnamefont{T.}~\bibnamefont{{Alexander}}},
  \bibinfo{journal}{\apj} \textbf{\bibinfo{volume}{645}}, \bibinfo{pages}{1152}
  (\bibinfo{year}{2006}), \eprint{astro-ph/0601161}.

\bibitem[{\citenamefont{{Merritt} et~al.}(2011)\citenamefont{{Merritt},
  {Alexander}, {Mikkola}, and {Will}}}]{mamw2}
\bibinfo{author}{\bibfnamefont{D.}~\bibnamefont{{Merritt}}},
  \bibinfo{author}{\bibfnamefont{T.}~\bibnamefont{{Alexander}}},
  \bibinfo{author}{\bibfnamefont{S.}~\bibnamefont{{Mikkola}}},
  \bibnamefont{and} \bibinfo{author}{\bibfnamefont{C.~M.}
  \bibnamefont{{Will}}}, \bibinfo{journal}{\prd} \textbf{\bibinfo{volume}{84}},
  \bibinfo{eid}{044024} (\bibinfo{year}{2011}), \eprint{1102.3180}.

\bibitem[{\citenamefont{{Einstein} et~al.}(1938)\citenamefont{{Einstein},
  {Infeld}, and {Hoffmann}}}]{EIH}
\bibinfo{author}{\bibfnamefont{A.}~\bibnamefont{{Einstein}}},
  \bibinfo{author}{\bibfnamefont{L.}~\bibnamefont{{Infeld}}}, \bibnamefont{and}
  \bibinfo{author}{\bibfnamefont{B.}~\bibnamefont{{Hoffmann}}},
  \bibinfo{journal}{Ann. Math.} \textbf{\bibinfo{volume}{39}},
  \bibinfo{pages}{65} (\bibinfo{year}{1938}).

\bibitem[{\citenamefont{Lorentz and Droste}(1917)}]{LorentzDroste}
\bibinfo{author}{\bibfnamefont{H.~A.} \bibnamefont{Lorentz}} \bibnamefont{and}
  \bibinfo{author}{\bibfnamefont{J.}~\bibnamefont{Droste}},
  \bibinfo{journal}{Versl.\ K.\ Akad.\ Wetensch.\ Amsterdam}
  \textbf{\bibinfo{volume}{26}}, \bibinfo{pages}{392} (\bibinfo{year}{1917}).

\bibitem[{\citenamefont{{Ag{\'o}n} et~al.}(2011)\citenamefont{{Ag{\'o}n},
  {Pedraza}, and {Ramos-Caro}}}]{2011PhRvD..83l3007A}
\bibinfo{author}{\bibfnamefont{C.~A.} \bibnamefont{{Ag{\'o}n}}},
  \bibinfo{author}{\bibfnamefont{J.~F.} \bibnamefont{{Pedraza}}},
  \bibnamefont{and}
  \bibinfo{author}{\bibfnamefont{J.}~\bibnamefont{{Ramos-Caro}}},
  \bibinfo{journal}{\prd} \textbf{\bibinfo{volume}{83}}, \bibinfo{eid}{123007}
  (\bibinfo{year}{2011}), \eprint{1104.5262}.

\bibitem[{\citenamefont{{Ramos-Caro} et~al.}(2012)\citenamefont{{Ramos-Caro},
  {Ag{\'o}n}, and {Pedraza}}}]{2012PhRvD..86d3008R}
\bibinfo{author}{\bibfnamefont{J.}~\bibnamefont{{Ramos-Caro}}},
  \bibinfo{author}{\bibfnamefont{C.~A.} \bibnamefont{{Ag{\'o}n}}},
  \bibnamefont{and} \bibinfo{author}{\bibfnamefont{J.~F.}
  \bibnamefont{{Pedraza}}}, \bibinfo{journal}{\prd}
  \textbf{\bibinfo{volume}{86}}, \bibinfo{eid}{043008} (\bibinfo{year}{2012}),
  \eprint{1206.5804}.

\bibitem[{\citenamefont{{Miller} and {Hamilton}}(2002)}]{2002ApJ...576..894M}
\bibinfo{author}{\bibfnamefont{M.~C.} \bibnamefont{{Miller}}} \bibnamefont{and}
  \bibinfo{author}{\bibfnamefont{D.~P.} \bibnamefont{{Hamilton}}},
  \bibinfo{journal}{\apj} \textbf{\bibinfo{volume}{576}}, \bibinfo{pages}{894}
  (\bibinfo{year}{2002}), \eprint{astro-ph/0202298}.

\bibitem[{\citenamefont{{Blaes} et~al.}(2002)\citenamefont{{Blaes}, {Lee}, and
  {Socrates}}}]{2002ApJ...578..775B}
\bibinfo{author}{\bibfnamefont{O.}~\bibnamefont{{Blaes}}},
  \bibinfo{author}{\bibfnamefont{M.~H.} \bibnamefont{{Lee}}}, \bibnamefont{and}
  \bibinfo{author}{\bibfnamefont{A.}~\bibnamefont{{Socrates}}},
  \bibinfo{journal}{\apj} \textbf{\bibinfo{volume}{578}}, \bibinfo{pages}{775}
  (\bibinfo{year}{2002}), \eprint{astro-ph/0203370}.

\bibitem[{\citenamefont{{Wen}}(2003)}]{2003ApJ...598..419W}
\bibinfo{author}{\bibfnamefont{L.}~\bibnamefont{{Wen}}},
  \bibinfo{journal}{\apj} \textbf{\bibinfo{volume}{598}}, \bibinfo{pages}{419}
  (\bibinfo{year}{2003}), \eprint{astro-ph/0211492}.

\bibitem[{\citenamefont{{Migaszewski} and
  {Go{\'z}dziewski}}(2011)}]{2011MNRAS.411..565M}
\bibinfo{author}{\bibfnamefont{C.}~\bibnamefont{{Migaszewski}}}
  \bibnamefont{and}
  \bibinfo{author}{\bibfnamefont{K.}~\bibnamefont{{Go{\'z}dziewski}}},
  \bibinfo{journal}{\mnras} \textbf{\bibinfo{volume}{411}},
  \bibinfo{pages}{565} (\bibinfo{year}{2011}), \eprint{1006.5961}.

\bibitem[{\citenamefont{{Seto}}(2013)}]{2013PhRvL.111f1106S}
\bibinfo{author}{\bibfnamefont{N.}~\bibnamefont{{Seto}}},
  \bibinfo{journal}{\prl} \textbf{\bibinfo{volume}{111}}, \bibinfo{eid}{061106}
  (\bibinfo{year}{2013}), \eprint{1304.5151}.

\bibitem[{\citenamefont{{Antognini} et~al.}(2013)\citenamefont{{Antognini},
  {Shappee}, {Thompson}, and {Amaro-Seoane}}}]{2013arXiv1308.5682A}
\bibinfo{author}{\bibfnamefont{J.~M.} \bibnamefont{{Antognini}}},
  \bibinfo{author}{\bibfnamefont{B.~J.} \bibnamefont{{Shappee}}},
  \bibinfo{author}{\bibfnamefont{T.~A.} \bibnamefont{{Thompson}}},
  \bibnamefont{and}
  \bibinfo{author}{\bibfnamefont{P.}~\bibnamefont{{Amaro-Seoane}}},
  \bibinfo{journal}{ArXiv e-prints}  (\bibinfo{year}{2013}),
  \eprint{1308.5682}.

\bibitem[{\citenamefont{{Naoz} et~al.}(2013)\citenamefont{{Naoz}, {Kocsis},
  {Loeb}, and {Yunes}}}]{2013ApJ...773..187N}
\bibinfo{author}{\bibfnamefont{S.}~\bibnamefont{{Naoz}}},
  \bibinfo{author}{\bibfnamefont{B.}~\bibnamefont{{Kocsis}}},
  \bibinfo{author}{\bibfnamefont{A.}~\bibnamefont{{Loeb}}}, \bibnamefont{and}
  \bibinfo{author}{\bibfnamefont{N.}~\bibnamefont{{Yunes}}},
  \bibinfo{journal}{\apj} \textbf{\bibinfo{volume}{773}}, \bibinfo{eid}{187}
  (\bibinfo{year}{2013}), \eprint{1206.4316}.

\bibitem[{\citenamefont{{Galaviz} and
  {Br{\"u}gmann}}(2011)}]{2011PhRvD..83h4013G}
\bibinfo{author}{\bibfnamefont{P.}~\bibnamefont{{Galaviz}}} \bibnamefont{and}
  \bibinfo{author}{\bibfnamefont{B.}~\bibnamefont{{Br{\"u}gmann}}},
  \bibinfo{journal}{\prd} \textbf{\bibinfo{volume}{83}}, \bibinfo{eid}{084013}
  (\bibinfo{year}{2011}), \eprint{1012.4423}.

\bibitem[{\citenamefont{{Galaviz}}(2011)}]{2011PhRvD..84j4038G}
\bibinfo{author}{\bibfnamefont{P.}~\bibnamefont{{Galaviz}}},
  \bibinfo{journal}{\prd} \textbf{\bibinfo{volume}{84}}, \bibinfo{eid}{104038}
  (\bibinfo{year}{2011}), \eprint{1108.4485}.

\bibitem[{\citenamefont{{Soffel} et~al.}(1988)\citenamefont{{Soffel}, {Wirrer},
  {Schastok}, {Ruder}, and {Schneider}}}]{1988CeMec..42...81S}
\bibinfo{author}{\bibfnamefont{M.}~\bibnamefont{{Soffel}}},
  \bibinfo{author}{\bibfnamefont{R.}~\bibnamefont{{Wirrer}}},
  \bibinfo{author}{\bibfnamefont{J.}~\bibnamefont{{Schastok}}},
  \bibinfo{author}{\bibfnamefont{H.}~\bibnamefont{{Ruder}}}, \bibnamefont{and}
  \bibinfo{author}{\bibfnamefont{M.}~\bibnamefont{{Schneider}}},
  \bibinfo{journal}{Celestial Mechanics} \textbf{\bibinfo{volume}{42}},
  \bibinfo{pages}{81} (\bibinfo{year}{1988}).

\bibitem[{\citenamefont{{Brumberg}}(1991)}]{1991ercm.book.....B}
\bibinfo{author}{\bibfnamefont{V.~A.} \bibnamefont{{Brumberg}}},
  \emph{\bibinfo{title}{{Essential relativistic celestial mechanics}}}
  (\bibinfo{publisher}{Adam Hilger}, \bibinfo{address}{Bristol},
  \bibinfo{year}{1991}), \bibinfo{note}{pp. 112 --115}.

\bibitem[{\citenamefont{{Pati} and {Will}}(2002)}]{patiwill2}
\bibinfo{author}{\bibfnamefont{M.~E.} \bibnamefont{{Pati}}} \bibnamefont{and}
  \bibinfo{author}{\bibfnamefont{C.~M.} \bibnamefont{{Will}}},
  \bibinfo{journal}{\prd} \textbf{\bibinfo{volume}{65}}, \bibinfo{eid}{104008}
  (\bibinfo{year}{2002}), \eprint{gr-qc/0201001}.

\bibitem[{\citenamefont{{Blanchet}}(2006)}]{lucLRR}
\bibinfo{author}{\bibfnamefont{L.}~\bibnamefont{{Blanchet}}},
  \bibinfo{journal}{Living Rev. Relativ.} \textbf{\bibinfo{volume}{9}},
  \bibinfo{pages}{4} (\bibinfo{year}{2006}).

\bibitem[{\citenamefont{{Saha} and {Tremaine}}(1994)}]{1994AJ....108.1962S}
\bibinfo{author}{\bibfnamefont{P.}~\bibnamefont{{Saha}}} \bibnamefont{and}
  \bibinfo{author}{\bibfnamefont{S.}~\bibnamefont{{Tremaine}}},
  \bibinfo{journal}{\aj} \textbf{\bibinfo{volume}{108}}, \bibinfo{pages}{1962}
  (\bibinfo{year}{1994}), \eprint{astro-ph/9403057}.

\end{thebibliography}

\end{document}